\documentclass[aps,prd,superscriptaddress,nofootinbib,eqsecnum,twocolumn]{revtex4}

\pdfoutput=1

\usepackage{amsfonts}
\usepackage{amsmath}
\usepackage{amssymb}
\usepackage{graphicx,color}
\usepackage{float}
\usepackage{hyperref}
\usepackage{subfigure}
\usepackage{mathtools,slashed}

\begin{document}

\title{Quantum chromodynamics axion in a hot and magnetized medium}

\author{Aritra Bandyopadhyay}
\affiliation{Departamento de Fisica, Universidade Federal de Santa Maria,
  97105-900 Santa Maria, RS,  Brazil}

\author{Ricardo L. S. Farias}
\affiliation{Departamento de Fisica, Universidade Federal de Santa Maria,
  97105-900 Santa Maria, RS,  Brazil}

\author{Bruno S. Lopes}
\affiliation{Departamento de Fisica, Universidade Federal de Santa Maria,
  97105-900 Santa Maria, RS,  Brazil}

\author{Rudnei O. Ramos}
\affiliation{Departamento de F\'{\i}sica Te\'orica, Universidade do
  Estado do Rio de Janeiro, 20550-013 Rio de Janeiro, RJ, Brazil}


\begin{abstract}

It is analyzed the effects of a hot and magnetized medium on the axion
mass, self-coupling and topological susceptibility when in presence of
an anisotropic external magnetic field along the $z$-direction, within
the Nambu--Jona-Lasinio effective model for quantum
chromodynamics. The effects of both Magnetic Catalysis and Inverse
Magnetic Catalysis are explicitly taken into account through
appropriate matching of parameters with those from lattice Monte-Carlo
numerical simulations. It is also analyzed the dependence of the
results with respect to different model parametrizations in the
context of the Nambu--Jona-Lasinio model.
 
\end{abstract}


\maketitle


\section{Introduction}
\label{sec1}

The axion is a pseudo Nambu-Goldstone boson of a spontaneously broken
global Abelian symmetry~\cite{Peccei:1977hh}.  The axion is considered
to be the most elegant and robust solution to the absence of the
charge and parity (CP) violating  effects (also dubbed the strong CP
problem) in quantum chromodynamics
(QCD)~\cite{Weinberg:1977ma,Wilczek:1977pj}.  It has also been
considered as a prime candidate for cold dark matter given that they
are very weakly coupled to  baryonic matter in general, besides of
possibly being extremely light (see, e.g., Ref.~\cite{Marsh:2015xka} for
a recent review and references therein).

The importance of the QCD axion as a solution to the strong CP problem and
its potential in explaining the dark matter abundance in the universe
makes it one of the most sought out prospects beyond the particle physics 
Standard Model. Recent studies also show that axions can thermalize and form a
Bose-Einstein condensate~\cite{Sikivie:2009qn,Chavanis:2017loo} which
in turn indicates the relevance of finite temperature extension of the
axion properties. On the other hand, the coupling of QCD axions with
an external electromagnetic field was first exploited long ago
in Ref.~\cite{Sikivie:1983ip} to make the axion experimentally
detectable. Since then various other experimental techniques involving
the axion-photon coupling have been
proposed~\cite{Sikivie:1985yu,VanBibber:1987rq,DePanfilis:1987dk} and
results of some such experiments are also
available~\cite{Hagmann:1990tj,Sikivie:1993jm}. 

Axions have also been associated  with the anomalous stellar cooling 
problem~\cite{Raffelt:2006cw,Giannotti:2015kwo}, including neutron stars~\cite{as1,as2,as3}.  
This is because axions could be produced in hot astrophysical plasmas and, thus, could
take part of the energy transport in stellar objects. This is in
particular a motivation for studying the properties of the QCD axion
when in presence of an environment that accounts not only thermal
effects (which can also be of relevance to the physics of these
particles in the early universe), but also density (chemical
potential) and external magnetic fields (relevant also for the physics
of compact stellar objects, like neutron stars).

As a pseudo Goldstone boson, the QCD axion acquires a mass, $m_a$,
from the QCD chiral symmetry breaking and this mass is typically of
$\mathcal{O}(m_\pi f_\pi/f_a)$~\cite{Weinberg:1977ma}, where $m_\pi$
is the pion mass, $f_\pi$ is the pion decay constant and $f_a$ is the
axion decay constant, which is proportional to the Peccei-Quinn
symmetry breaking scale. We can also define an effective
self-coupling, $\lambda_a$, to the axion field.  {}Furthermore, there
are  derivative couplings that can also be present in a system
involving axions.  The axion mass and the couplings (including
the self-coupling) are all controlled by the scale $f_a$.  Though exact
determination of the value of $f_a$ is not yet achievable, present
astrophysical constraints suggest the value for $f_a$ to be in between
$10^8$ GeV and $10^{18}$
GeV~\cite{Raffelt:2006cw,Arvanitaki:2010sy,Arvanitaki:2014wva}.  This
makes the couplings associated with axion field extremely small,
making it experimentally challenging to be probed (see,
e.g., Ref.~\cite{Brun:2019lyf}  and references therein for some of the
experimental proposals looking for axions).

The extension of axion studies in relatively higher temperatures can
be done perturbatively, using, e.g., the dilute instanton gas
approximation~\cite{Turner:1985si}. But around and below the pseudo 
critical temperature, $T_c\sim 170$ MeV, which are particularly important in
relation with the chiral symmetry breaking, there are no reliable
perturbative techniques. Non-perturbative methods making use of
effective models come into play in this regime of study along with
the first principle calculations of Lattice QCD. In this particular
work, we have chosen to deal with one of such effective model, the
Nambu--Jona-Lasinio (NJL) model~\cite{Klevansky:1992qe} to study the
QCD axions in a hot and magnetized medium (For a recent study of axion 
within a chiral effective Lagrangian model, see Ref~\cite{Landini:2019eck}). Beside its relevance to
spontaneous CP
violation~\cite{Frank:2003ve,Boomsma:2009eh,Boomsma:2009yk,Chatterjee:2014csa,Fukushima:2001hr},
the advantage of using a two-flavor NJL model to study a system of
axions is that being a quark model, it incorporates the effects of
axions via the effective (`t Hooft determinant) interaction~\cite{tHooft:1976snw,tHooft:1986ooh}. 
As for the electromagnetic
coupling of the axions, for this work we will only consider an
anisotropic external magnetic field along the $z$-direction. In the
future we plan to extend the present study for other general systems, where both
electric and magnetic fields can be present in the system. 
As an additional remark, since the energy scales 
we are working with are much smaller than the axion symmetry breaking scale $f_a$, 
we can consider the axion field to be in its (constant) vacuum expectation value,
so it behaves like a CP violating term added to the QCD action. Thus, in this sense,
our derivations can be performed in a way similar like in previous 
studies~\cite{Frank:2003ve,Boomsma:2009eh,Boomsma:2009yk,Chatterjee:2014csa,Fukushima:2001hr}.

This paper is organized as follows. In Sec.~\ref{sec2} we discuss the
formalism of the NJL model when incorporating the axion field. We also
explain the derivation of the thermodynamic potential for the model in
the cases without and with an external anisotropic magnetic field.
The way the parameters are fixed is also discussed. In Sec.~\ref{sec3}
we present our results concerning the analysis of the effects of
temperature and magnetic field on the axion mass, coupling constant
and also on the axion topological susceptibility, which is an important observable
derivable for example from lattice QCD results. {}Finally, in Sec.~\ref{sec4}
we have our conclusions. 


\section{Formalism}
\label{sec2}

We start with a brief review of the QCD axion and how its associated
field is included in the NJL model quasi-particle description of quark
matter.  Then, we will discuss the incorporation of both temperature
and an external anisotropic magnetic field in the derivation of the
thermodynamic potential for the model. While parametrizing the model,
focus will also be given on the recently discovered phenomena of
Magnetic Catalysis (MC)~\cite{Gusynin:1995nb} and Inverse Magnetic Catalysis 
(IMC)~\cite{Bali:2011qj,Bali:2012zg} on the
thermodynamic potential when in presence of a magnetized medium.

\subsection{The axion contribution}

The study of CP violation within strongly interacting matter has been a subject of
extensive scrutiny over the years~\cite{Frank:2003ve,Boomsma:2009eh,Boomsma:2009yk,Chatterjee:2014csa,Fukushima:2001hr,Fraga:2009vy,Mizher:2008hf,Mizher:2008dj,Fraga:2008um}. 
It is a well known fact that within
the regime of strong interaction instanton contributions can lead to
CP violation~\cite{tHooft:1976snw,tHooft:1986ooh}. In this kind of
scenarios where gauge field configurations have nontrivial topologies,
the QCD Lagrangian density generally contains an extra $\theta$-term, 
\begin{eqnarray}
\mathcal{L}_\theta = \frac{\theta g^2}{32\pi^2}
\mathcal{F}\bar{\mathcal{F}},
\label{ltheta}
\end{eqnarray}
where $g$ is the coupling for the strong interaction and 
$\mathcal{F}$ and $\bar{\mathcal{F}}$ are the gluonic field
strength tensor and its dual respectively. The real parameter $\theta$
defines the choice of vacuum from infinite possibilities and its value
subsequently dictates whether the corresponding theory is CP symmetric
or not. As can be seen straightforwardly from Eq.~(\ref{ltheta}), only
for $\theta = 0$ (mod $\pi$), the QCD Lagrangian is CP
conserving. Again various experimental studies on pseudoscalar mass
ratios~\cite{Kawarabayashi:1980uh}, electrical dipole
moments~\cite{Baluni:1978rf,Crewther:1979pi,Baker:2006ts,Afach:2015sja}
as well as Lattice QCD
calculations~\cite{Guo:2015tla,Bhattacharya:2015esa} conclude that the
value of this real angular parameter $\theta$ is very close to 0 in
nature.  A simple and elegant way to explain why $\theta$ should be so
small or null is giving $\theta$ a dynamical character, elevating it
to a field, the axion, such as to have a vanishing vacuum expectation
value~\cite{Peccei:1977hh}.  The axion field $a$ is the canonically
normalized dynamical $\theta$, $\theta(x) = a(x)/f_a$, where $f_a$ is
the axion decay constant. Its only non-derivative coupling is to the
QCD topological charge and it is suppressed by the scale $f_a$. The
interaction Lagrangian density in Eq.~(\ref{ltheta}) can now be 
written as 
\begin{eqnarray}
\mathcal{L}_a= \frac{g^2}{32\pi^2}\frac{a}{f_a}
\mathcal{F}\bar{\mathcal{F}}.
\label{laga1}
\end{eqnarray}
Equation~(\ref{laga1}) can be effectively represented as an
interaction of the QCD axion field $a$ with the quarks by performing a
chiral rotation~\cite{Buballa:2003qv} of the quark fields by the angle
$a/f_a$, which yields~\cite{tHooft:1976snw,tHooft:1986ooh}
\begin{eqnarray}
\mathcal{L}_a = 8G_2
\left[e^{i\frac{a}{f_a}}\text{det}(\psi_R\psi_L)+e^{-i\frac{a}{f_a}}
  \text{det}(\psi_L\psi_R)\right],
\label{laga2}
\end{eqnarray}
where $\psi_L$ and $\psi_R$ are the left- and right-handed components
of the quark wave function $\psi$ and $G_2$ is a coupling constant.

\subsection{Thermodynamic potential for the axion background field within 
the NJL model}
\label{subsec2a}

The effective Lagrangian density for the isospin symmetric two-flavor
NJL model with the CP violating term is given by
\begin{eqnarray}
\mathcal{L} = \bar{\psi} (i\gamma^\mu\partial_\mu-m_0)\psi +
\mathcal{L}_q + \mathcal{L}_a,
\end{eqnarray}
where $\psi$ depicts the fermionic (quark) fields and $m_0$ is the
current quark mass. The fermionic interaction part of the Lagrangian
density is given by 
\begin{eqnarray}
\mathcal{L}_q = G_1 \left[(\bar{\psi}\psi)^2+(\bar{\psi}\tau_k i
  \gamma_5\psi)^2\right], 
\label{lag_q1}
\end{eqnarray}
where $\tau_k$, $k=0,1,2,3$, represents the unit matrix (for $k=0$)
and the Pauli matrices (for $k=1,2,3$) and $G_1$ is the coupling
constant associated with the fermionic interaction.  Lastly, the
symmetry breaking 't Hooft determinant interaction term ${\cal L}_a$
is given by Eq.~(\ref{laga2}).  The axion contribution
Eq.~(\ref{laga2}) effectively breaks down the global U$(2)_V$ symmetry
into SU$(2)_V \times$ U$(1)_B$. Often, in recent studies involving flavor mixing within 
the NJL model and its extensions, 
the couplings $G_1$
and $G_2$ are also taken to be equal~\cite{Frank:2003ve,Buballa:2003qv}. 
In the present work we choose $G_1 =(1-c) G_s$ and $G_2 = c\, G_s$, following
Refs.~\cite{Frank:2003ve,Boomsma:2009eh,Lu:2018ukl}, which assumes the
connection of both the coupling constants with $G_s$, the standard
scalar channel coupling constant for the NJL model. The parameter $c$
connecting the two couplings determines the strength of the axion
interaction and we will discuss more about its value in
Section~\ref{subsec2c}. 

Next, for the derivation of the thermodynamic potential, we use the
usual mean-field approximation~\cite{Klevansky:1992qe}, where the scalar 
and pseudoscalar fields are  replaced by their corresponding mean-field 
values, or condensates,  
\begin{eqnarray}
&&\bar{\psi}\psi \rightarrow \langle \bar{\psi}\psi \rangle = \sigma,
  \\  &&\bar{\psi}i \gamma_5\psi \rightarrow \langle \bar{\psi}i
  \gamma_5\psi \rangle = \eta,
\end{eqnarray} 
with $\sigma$ and $\eta$ representing the chiral and pseudoscalar
condensates, respectively. Note that in the following we will be
considering only the isospin symmetric case, hence, only $\langle
\bar{\psi}\psi \rangle$ and $\langle \bar{\psi}i \gamma_5\psi \rangle$
survive from Eq.~(\ref{lag_q1}). Usually, for $\theta=0$, or in the
present case, for $a=0$, $\langle \bar{\psi}i \gamma_5\psi \rangle$
also vanishes. Considering a nonvanishing $\langle \bar{\psi}i
\gamma_5\psi \rangle$ emphasizes explicitly the fact that the axion
field couples to the axial current. Hence, in terms of $\sigma$ and
$\eta$, the thermodynamic potential for the QCD axion within the NJL
model, following for instance Ref.~\cite{Lu:2018ukl}, is given by
\begin{eqnarray}
\Omega &=& \Omega_q + G_1 (\eta^2 +\sigma^2) - G_2 (\eta^2 -
\sigma^2)\cos \frac{a}{f_a} \nonumber\\  &&- 2G_2 \sigma \eta \sin
\frac{a}{f_a},
\label{omega_axion}
\end{eqnarray}
where the quark contribution $\Omega_q$ is given
by~\cite{Klevansky:1992qe,Buballa:2003qv}
\begin{eqnarray}
\Omega_q = -8N_c \int \frac{d^3p}{(2\pi)^3}
\left[\frac{E_p}{2}+T\ln \left(1+e^{-E_p/T}\right)\right],
\label{omegaq}
\end{eqnarray}
where $N_c=3$ is the number of colors, $E_p = \sqrt{p^2+M^2}$ and  
\begin{eqnarray}
M=\sqrt{(m_0+\alpha_0)^2+\beta_0^2},
\label{eff_mass}
\end{eqnarray}
and $\alpha_0$ and $\beta_0$ in the effective mass $M$, Eq.~(\ref{eff_mass}),
can be written in terms of the axion background field $a$ and the
condensates $\sigma$ and $\eta$ as~\cite{Lu:2018ukl}
\begin{eqnarray}
\!\!\!\!\!\alpha_0 &=& -2 \left(G_1 + G_2 \cos \frac{a}{f_a}\right)\sigma +
2G_2\eta\sin\frac{a}{f_a},
\\ 
\!\!\!\!\!\beta_0 &=& -2 \left(G_1 - G_2 \cos
\frac{a}{f_a}\right)\eta + 2G_2\sigma\sin\frac{a}{f_a}.
\label{alpha_beta}
\end{eqnarray}
The momentum integral in Eq.~(\ref{omegaq}) consists of a vacuum part
(the first term inside the argument of the integral) and a medium part 
(the second term inside the argument of the integral). The vacuum contribution
is ultraviolet (UV) divergent and usually taken care of by a sharp
cut-off regularization procedure, i.e., with a finite three-momentum
upper cut-off $\Lambda$.  It can be checked in a straightforward way
that the vanishing axion field limit, i.e., setting $a\to 0$ in
Eq.~(\ref{omega_axion}), reproduces the usual NJL thermodynamic
potential~\cite{Klevansky:1992qe}.

{}From the thermodynamic potential $\Omega$, Eq.~(\ref{omega_axion}),
we can now find the physical values for the condensates  $\sigma$ and
$\eta$ by solving the appropriate gap equations,
\begin{eqnarray}
&&\frac{\partial \Omega (\sigma, \eta, \frac{a}{f_a})}{\partial
    \sigma}\Bigg|_{\sigma=\sigma_0 \atop \eta=\eta_0} = 0, \nonumber
  \\  &&\frac{\partial \Omega (\sigma, \eta, \frac{a}{f_a})}{\partial
    \eta}\Bigg|_{\sigma=\sigma_0\atop \eta=\eta_0} = 0,
\label{gap_eqn}
\end{eqnarray}
which also depend on the value of the axion background field through the ratio
$a/f_a$. The effective thermodynamic potential for the QCD axion in a
hot medium and within NJL model is then given by
\begin{eqnarray}
\Omega_T (a,T) = \Omega \left[\sigma_0(a, T), \eta_0(a, T), a,
  T\right].
\end{eqnarray}
Since in the present study the axion is treated as a background
field, the axion mass $m_a$ is simply defined by the second
derivative of the effective potential at vanishing axion field, i.e.,
\begin{eqnarray}
m_a^2 = \frac{d^2\Omega_T (a, T)}{da^2} \Bigg|_{a=0} = \frac{\chi_{\rm
    top}}{f_a^2},
\label{mass_axion}
\end{eqnarray}
where $\chi_{	\rm top }$ is the topological susceptibility, 
which is independent of the scale $f_a$, as it is evident from
Eq.~(\ref{mass_axion}).  Similarly, the axion self-coupling
$\lambda_a$ is defined as the fourth derivative of the effective
potential at vanishing axion field limit,
\begin{eqnarray}
\lambda_a = \frac{d^4\Omega_T (a, T)}{da^4} \Bigg|_{a=0}.
\label{selfc_axion}
\end{eqnarray}

\subsection{Adding an external magnetic field}
\label{subsec2b}

We now turn to the modification of the thermodynamic potential when in
the presence of an external magnetic field.  The effective Lagrangian
density of the QCD axion in the presence of an external
electromagnetic (EM) field, within the two-flavor NJL model and at leading
order in $1/f_a$, can be written as
\begin{eqnarray}
\mathcal{L}_{\textrm{EM}} &=& \bar{\psi} (i\gamma^\mu D_\mu-m_0)\psi +
\mathcal{L}_q + \mathcal{L}_a \nonumber\\ &&- \frac{1}{4}F^{\mu \nu}
F_{\mu \nu} + g_{\gamma a}\frac{a}{4}F^{\mu \nu} \tilde{F}_{\mu \nu},
\end{eqnarray}
where $D_\mu = \partial_\mu - i q A_\mu$, $q$ being the electric
charge and $A_\mu$ the EM gauge field. $F^{\mu \nu}$ is the field
strength tensor given by $F^{\mu \nu} = \partial^\mu A^\nu -
\partial^\nu A^\mu$ and $\tilde{F}_{\mu \nu}$ is its dual. 
The axion-photon coupling constant $g_{\gamma a}$ appearing in 
the interaction term is given by~\cite{diCortona:2015ldu}
\begin{eqnarray}
g_{\gamma a} = \frac{\alpha_{\rm em}}{2\pi}\frac{E}{N},
\end{eqnarray}
where $\alpha_{\rm em}$ is the EM fine-structure constant and $E/N$ is
the EM to color anomaly ratio (for example, with the value of $8/3$ in
light of the Grand Unification Theory (GUT) $SU(5)$ model~\cite{diCortona:2015ldu}). 

As mentioned in Sec.~\ref{sec1}, for our present study we consider an
external anisotropic magnetic field along the $z$-direction. In this
case, the axion-photon coupling term vanishes ($F^{\mu \nu}
\tilde{F}_{\mu \nu} \propto {\bf E}.{\bf B} \rightarrow 0$) and the 
effective Lagrangian density simplifies to 
\begin{eqnarray}
\!\!\!\!\! \mathcal{L}_{\textrm{B}}=\bar{\psi} (i\gamma^\mu D_\mu-m_0)\psi +
\mathcal{L}_q + \mathcal{L}_a - \frac{1}{4}F^{\mu \nu} F_{\mu \nu}.
\end{eqnarray}
In the presence of the anisotropic external magnetic field along the
$z$-direction, the transverse plane in the momentum space gets
quantized and the dispersion relation for the quarks modifies to~\cite{Ebert:1999ht} 
\begin{eqnarray}
E_p'(B) = \left[ M^2 + p_z^2 + (2n+1-s) q_fB\right]^{1/2},
\end{eqnarray}
where $n$ and $s$ represent the Landau levels and the spin states,
respectively, and $q_f$ is the absolute charge of the fermion with
flavor\footnote{We would like to note here that usually in literature 
it is a common practice to use $|q_fB|$ where $q_f$ is expressed in terms of $e$ and 
it also carries the sign of the charge.}  $f\equiv u\, d$. Using $2l=(2n+1-s)$, the above relation can also be
written as 
\begin{eqnarray}
E_p(B) = \left[ M^2 + p_z^2 + 2 l q_f B\right]^{1/2},
\end{eqnarray}
$l$ being the redefined index for the Landau levels. As we shall see,
this reorganization of variables produces the degeneracy factor
$(2-\delta_{l,0})$ in the expression, counting the spin states for all
except the lowest Landau level.  Thus, by incorporating the quantized
transverse momenta, the three momentum integral becomes
\begin{eqnarray}
&&\int \frac{d^3p}{(2\pi)^3}f(E_p) \rightarrow \frac{q_fB}{2\pi}
  \sum\limits_{n=0}^\infty \int\limits_{-\infty}^\infty
  \frac{dp_z}{2\pi} f(E_p'(B)) \nonumber\\ &&\rightarrow
  \frac{q_fB}{2\pi} \sum\limits_{l=0}^\infty (2-\delta_{l,0})
  \int\limits_{-\infty}^\infty \frac{dp_z}{2\pi}f(E_p(B)).
\end{eqnarray}
The quark part of the thermodynamic potential $\Omega_q$ now gets
modified in the presence of the magnetic field and becomes
\begin{eqnarray}
\Omega_q(B,T) &=& -\frac{N_c}{\pi^2} \sum\limits_{l,f}
q_fB(2-\delta_{l,0}) \int\limits_{-\infty}^\infty dp_z
\left\{ \frac{E_p(B)}{2} \right. 
  \nonumber\\ 
&& \left. +T \ln \left[ 1+e^{-E_p(B)/T}\right] \right\},
\end{eqnarray}
and the total thermodynamic potential for the present system can be
written as,
\begin{eqnarray}
\Omega &=& \Omega_q (B,T) + G_1 (\eta^2 +\sigma^2) - G_2 (\eta^2 -
\sigma^2)\cos \frac{a}{f_a} \nonumber\\  &&- 2G_2 \sigma \eta~ \sin
\frac{a}{f_a} .
\label{omegaB_1}
\end{eqnarray}
It is important to note here that in the presence of the external magnetic field, 
the condensates corresponding to the $u$ and $d$ quarks are not the same anymore 
due to the presence of the factor $q_f$ in $\Omega_q (B,T)$. Hence, in Eq.~(\ref{omegaB_1}), 
the new $\sigma$ and $\eta$ condensates are basically the average with respect to the quark flavors, i.e., 
$\sigma = (\sigma_u+\sigma_d)/2$ and $\eta = (\eta_u+\eta_d)/2$. 

{}For convenience, the thermodynamic potential can also be rearranged
in three separate parts using  the Magnetic Field Independent
Regularization (MFIR) procedure. MFIR was proposed in ~\cite{Ebert:1999ht,Vdovichenko:2000sa,Ebert:2003yk} 
and has been recently applied 
in several works~\cite{Menezes:2008qt,Menezes:2009uc,farias,Allen:2015paa,Duarte:2016pgi,Avancini:2015ady,Duarte:2015ppa,Avancini:2016fgq,Farias:2016gmy,Coppola:2017edn,Avancini:2017gck,diff_regul,mpi0BT}. Using this procedure the thermodynamic potential becomes
\begin{equation}
\Omega (B,T) = \Omega_V + \Omega_B (B) + \Omega_M
(T,B), \label{omegaB_final}
\end{equation}
where $\Omega_V$ corresponds to the vacuum part, $\Omega_B (B)$ a term
that carries the dependence on $B$ only and $\Omega_M (T,B)$ is the medium
part which is dependent on both $T$ and $B$. Each one of these terms
are given, respectively, as
\begin{eqnarray}
\Omega_V &=& G_1 (\eta^2 +\sigma^2) - G_2 (\eta^2 -
\sigma^2)\cos \frac{a}{f_a} \nonumber\\ && - 2G_2 \sigma \eta \sin
\frac{a}{f_a} - 4 N_c \int\limits_\Lambda \frac{d^3p}{(2\pi)^3} E_p,\\ 
\Omega_B &=&  - \frac{ N_c}{2\pi^2} \sum\limits_f
(q_fB)^2 
\nonumber\\  &&\times\left[\zeta'(-1,x_f)-\frac{1}{2}(x_f^2-x_f)\ln
  x_f+\frac{x_f^2}{4}\right],  \nonumber\\
  \\ 
 \Omega_M  &=& -
\frac{N_c}{2\pi^2}  \sum\limits_{l,f} (2-\delta_{l,0})~q_fB
\nonumber\\  &&\times\int\limits_{-\infty}^\infty dp_z~ T \ln
\left[ 1+e^{-E_p(B)/T} \right],
\end{eqnarray}
where $x_f=M^2/(2q_f B)$ and $\Lambda$ is the finite
three-momentum cut-off introduced in Section~\ref{subsec2a}.  The
first derivative of the Hurwitz zeta function, $\zeta'(-1,x_f)$, can be
written in a simplified form that helps to evaluate  the derivatives
numerically without hassle. It is done by differentiating and
integrating the function with respect to $x_f$, leading to
\begin{eqnarray}
\zeta'(-1,x_f) =
\zeta'(-1,0)\!+\!\frac{x_f^2}{2}\!-\!\frac{x_f}{2}[1+\ln(2\pi)] \!+\!
\psi^{(-2)}(x_f). \nonumber
\end{eqnarray}
Hence, we get for $\Omega_B$ the result
\begin{eqnarray}
\Omega_B &=&  - \frac{ N_c}{2\pi^2}  \sum\limits_f(q_fB)^2
\left\{ \frac{3x_f^2}{4}-\frac{x_f}{2}[1+\ln(2\pi)] \right. 
\nonumber 
\\ 
&& \left. +  \psi^{(-2)}(x_f)-\frac{1}{2}(x_f^2-x_f)\ln x_f\right\},
\end{eqnarray}
where $\psi^{(m)}(x_f)$ represents the $m$-th polygamma function 
and the $x_f$ independent term $\zeta'(-1,0)$ has been
neglected. Subsequently, using Eq.~(\ref{omegaB_final}) as the
thermodynamic potential for the case of axions in a hot and magnetized
medium within the NJL model, we can solve the gap equations
(\ref{gap_eqn})  to get the $T$ and $B$ dependent condensates
vis-a-vis effective potential and, hence, obtain the $T$ and $B$
dependent axion mass, the axion self-coupling and the topological
susceptibility from Eqs.~(\ref{mass_axion}) and (\ref{selfc_axion}). 

\subsection{Parametrization}
\label{subsec2c}

Before getting the results from the thermodynamic potential, we need
to discuss the choice of parametrization within the NJL model.  The
usual parameters for the two-flavor NJL model have been introduced in
subsections~\ref{subsec2a} and \ref{subsec2b}. There are various
parameter sets used in the literature over the years to describe the
system within the NJL model.  Prior to moving into the values of different
parameters used in the present study, we discuss here the complexities
of dealing with two interesting and recently discovered phenomena in a
hot and magnetized medium, the MC and IMC phenomena. Earlier studies
of such medium effects using the NJL model have shown that the
spontaneous chiral symmetry breaking gets enhanced when in the
presence of a strong constant magnetic field  through the generation
of a fermion dynamical mass~\cite{Gusynin:1995nb,Schramm:1991ex}. This
phenomenon is commonly known as the MC. The existence of the MC was
further solidified by many other effective model
studies~\cite{Boomsma:2009yk,Chatterjee:2011ry,Ferrari:2012yw,Ferreira:2013tba}. On
the other hand, different lattice QCD simulations results found that even
though the values of the light quark condensates increase at
temperatures distant from $T_c$, it decreases near
$T_c$~\cite{Bali:2011qj,Bali:2012zg}. This counterintuitive behaviour
was dubbed as IMC.  The dominance of sea contributions over the
valence contributions of the condensate around $T_c$ is one of the
probable reasons~\cite{Bruckmann:2013oba}  behind the IMC. After the
recognition of the presence of IMC through lattice QCD for both chiral
and deconfinement
transitions~\cite{Bruckmann:2013oba,Endrodi:2015oba}, several attempts
have been made to understand it through different effective QCD
models~\cite{farias,Farias:2016gmy,mpi0BT,Yu:2014xoa,Mao:2016fha,Providencia:2014txa,Ferreira:2014kpa,Ayala:2014iba,Ayala:2014gwa,Ferrer:2014qka,endrodiGB}. At
this point we want to emphasize that in the present work we will be
discussing about both the effects of MC and IMC and the way of
incorporating them in the axion thermodynamic potential, which is one
of the novel features related to the present work.

In the present study, to observe the parametrization dependence of the
various quantities evaluated in the vanishing magnetic field limit, we
have used three different parameter sets, given in
Table~\ref{parameter_sets}.
By using different parameter sets, it will enable us to quantify the
dependence of the results on them.

\begin{table}
\caption{Different parameter sets are listed from different
  references, which have been used in the present study. }
\begin{center}
\begin{tabular}{c|c|c}
\hline Sets & Input parameters & Output parameters \\ \hline I &
$f_\pi = 93$ MeV  & $\Lambda = 650$ MeV
\\ Ref.~\cite{farias,Farias:2016gmy}&  $m_\pi = 140$ MeV & $G_s =
2.122/\Lambda^2$ \\ & $\langle \bar{\psi}\psi
\rangle^{\frac{1}{3}}=-250$ MeV & $m_0 = 5.5$ MeV \\ \hline II &
$f_\pi = 92.6$ MeV  & $\Lambda = 590$ MeV
\\ Ref.~\cite{Frank:2003ve,Lu:2018ukl}&  $m_\pi = 140.2$ MeV & $G_s =
2.435/\Lambda^2$ \\ & $\langle \bar{\psi}\psi
\rangle^{\frac{1}{3}}=-241.5$ MeV & $m_0 = 6$ MeV \\ \hline III &
$f_\pi = 92.4$ MeV  & $\Lambda = 664.3$ MeV
\\ Ref.~\cite{Buballa:2003qv}&  $m_\pi = 135$ MeV & $G_s =
2.06/\Lambda^2$ \\ & $\langle \bar{\psi}\psi
\rangle^{\frac{1}{3}}=-250.8$ MeV & $m_0 = 5$ MeV \\ \hline
\end{tabular}
\end{center}
\label{parameter_sets}
\end{table}
 
{}For MC we have considered Set I from
Refs.~\cite{farias,Farias:2016gmy} with a fixed value of the coupling
constant $G_s$. To incorporate IMC in the model we have also chosen
the procedure taken in Ref.~\cite{Farias:2016gmy}, i.e., by keeping
the other parameters same as in Set I and using a four-fermion scalar
coupling constant that is dependent on both the temperature and
magnetic field. This was proposed for the case of the $SU(2)$ NJL model in
Ref.~\cite{Farias:2016gmy}, such that 
\begin{eqnarray}
G_s(eB,T) &=& b(eB)
\left\{ 1-\frac{1}{1+e^{\beta(eB)[T_a(eB)-T]}}\right\}\nonumber\\&&+s(eB),
\label{imc_fit}
\end{eqnarray}
where the aforementioned parameters $b, \beta, T_a$ and $s$ (see 
Table~\ref{imc_parameters}) are obtained by fitting the lattice data for the
average of the quark condensates~\cite{Bali:2012zg}.

\begin{table}
\caption{Values of different fitting parameters used in
  Eq.~(\ref{imc_fit}). }
\label{imc_parameters}
\begin{center}
\begin{tabular}{c c c c c}
\hline $eB$ & $b$ & $T_a$ & $s$ & $\beta$ \\ ~(GeV)~ & ~(GeV)~ &
~(GeV)~ & ~(GeV)~ & ~(GeV)~ \\ \hline 0.0 & 0.9  & 0.168 & 3.731 &
40.0 \\ 0.2 & 1.226  & 0.168 & 3.262 & 34.117 \\ 0.4 & 1.769  & 0.169
& 2.294 & 22.988 \\ \hline
\end{tabular}
\end{center}
\end{table}

At this point we would also like to mention that scarcity of lattice
data at lower values of temperature eventually requires this
procedure to be performed with the fitting at $T>110$ MeV and to extrapolate
through the region of lower temperatures, which can give rise to
ambiguities over the value of the coupling constant $G_s(eB,T=0)$. This
dilemma can be avoided all together following, e.g., the procedure explained in
Ref.~\cite{Avancini:2016fgq}, where the authors have generated a
coupling constant $G_0(eB)$ for T=0, as a good fit to the lattice
simulations using selected values of $eB$ from $0$ to $1$ GeV$^2$,
given as
\begin{eqnarray}
G_0(eB) = \alpha + \beta e^{-\gamma (eB)^2},
\label{T0_Gfit}
\end{eqnarray}
with values of $\alpha, \beta$ and $\gamma$ given, respectively, as
$1.44373$ GeV$^{-2}$, $3.06$ GeV$^{-2}$ and $1.31$ GeV$^{-4}$. Using
this magnetic field dependent coupling $G_0(eB)$ in our present 
study\footnote{We would like to note here that $G_0(eB=0)$ 
comes out to be $4.50373$ GeV$^{-2}$, which is different from the value of $G_s$ shown
in the parameter Set I. 
This discrepancy slightly affects the values of the other parameters, e.g., 
$f_\pi$ becomes $91.22$ MeV instead of $93$ MeV. {}For our results using 
$G_0(eB)$, we have neglected these changes.},
we will separately show the behavior of the magnetic field dependence
of the topological susceptibility $\chi_{	\rm{top}}$ at $T=0$.

\begin{figure*}[!]
\begin{center}
\subfigure[Case Set I]{\includegraphics[width=8cm]{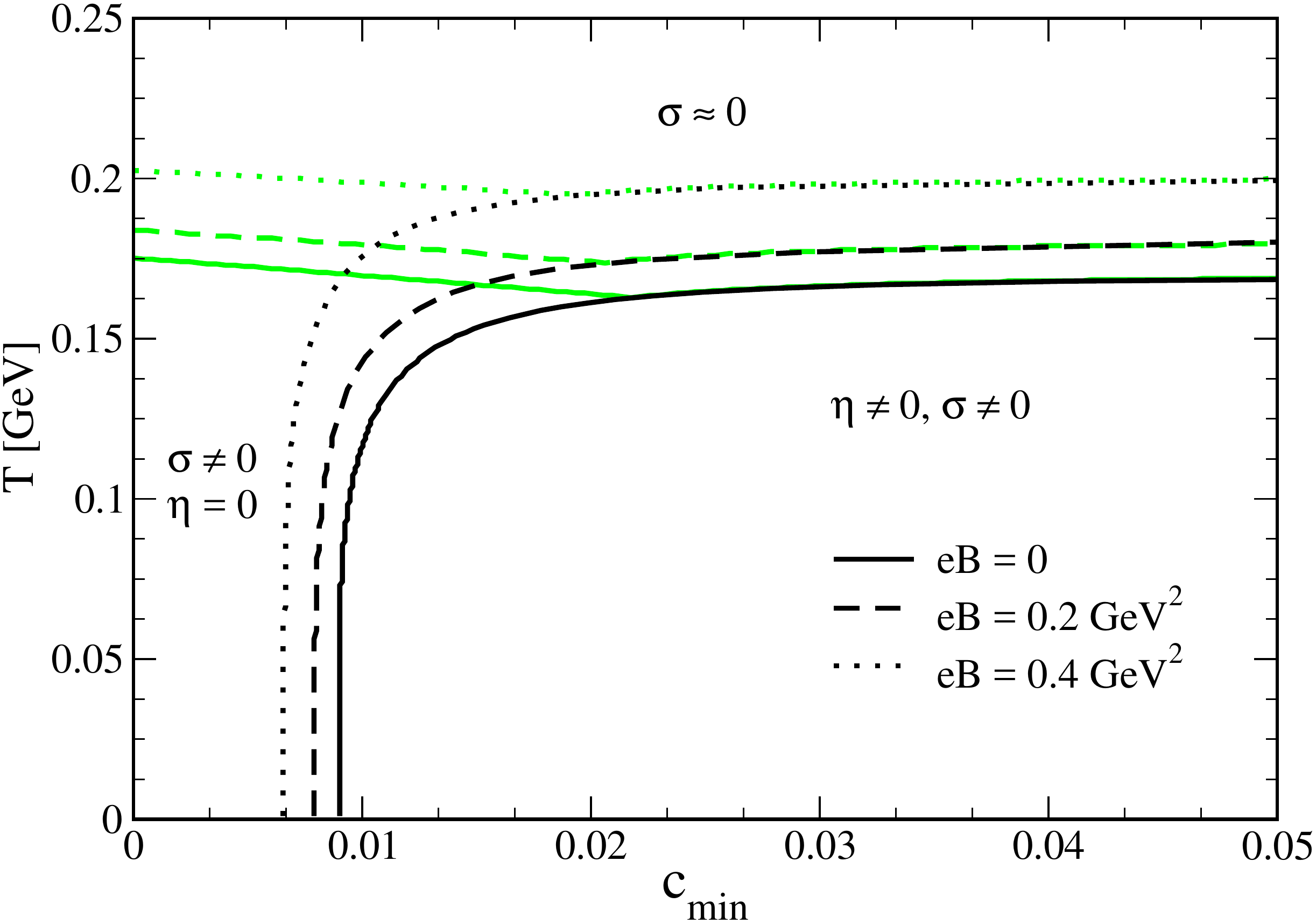}}
\subfigure[Case Set II]{\includegraphics[width=8cm]{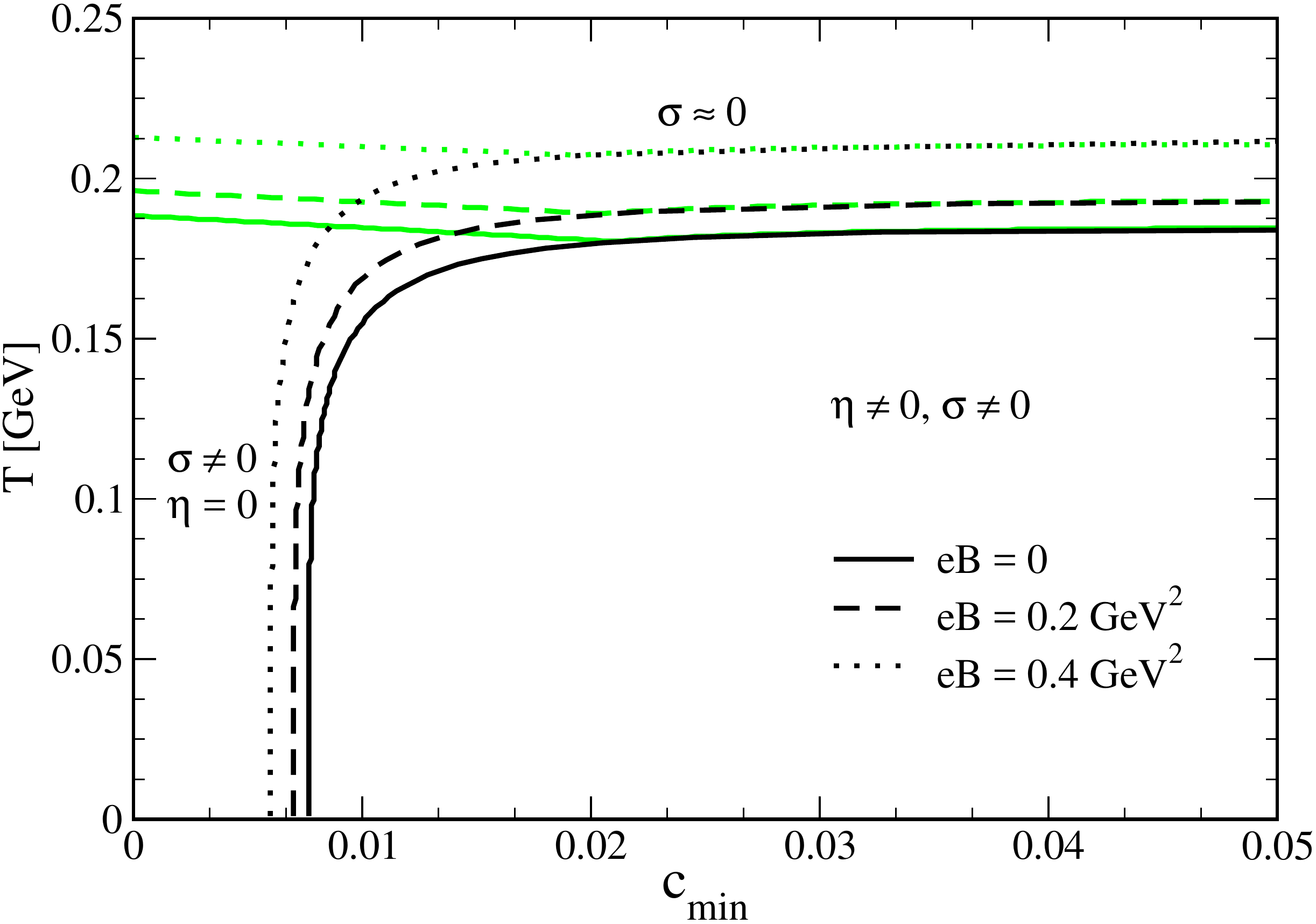}}
\subfigure[Case Set III]{\includegraphics[width=8cm]{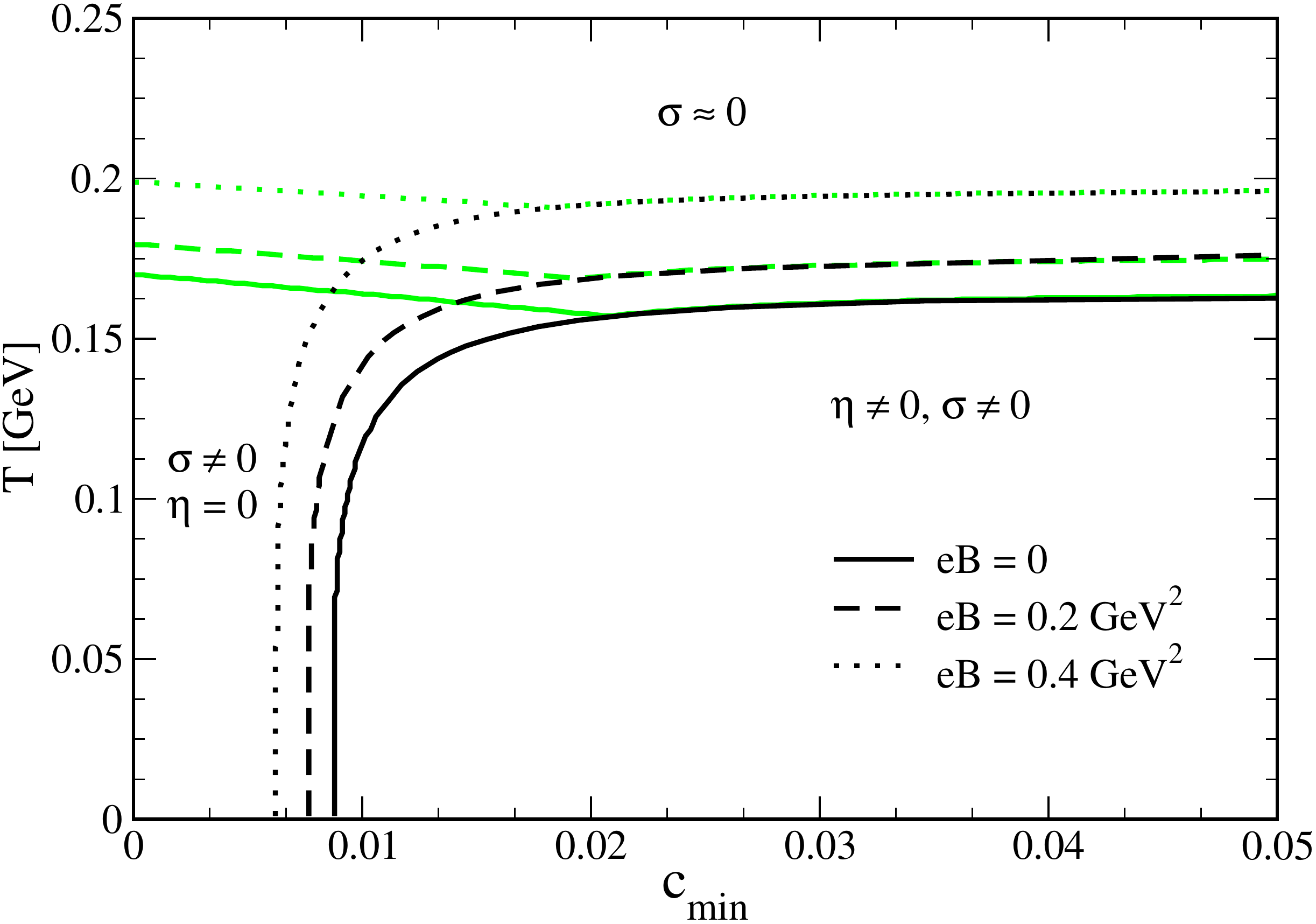}}
\subfigure[Case with $G_s(eB,T)$]{\includegraphics[width=8cm]{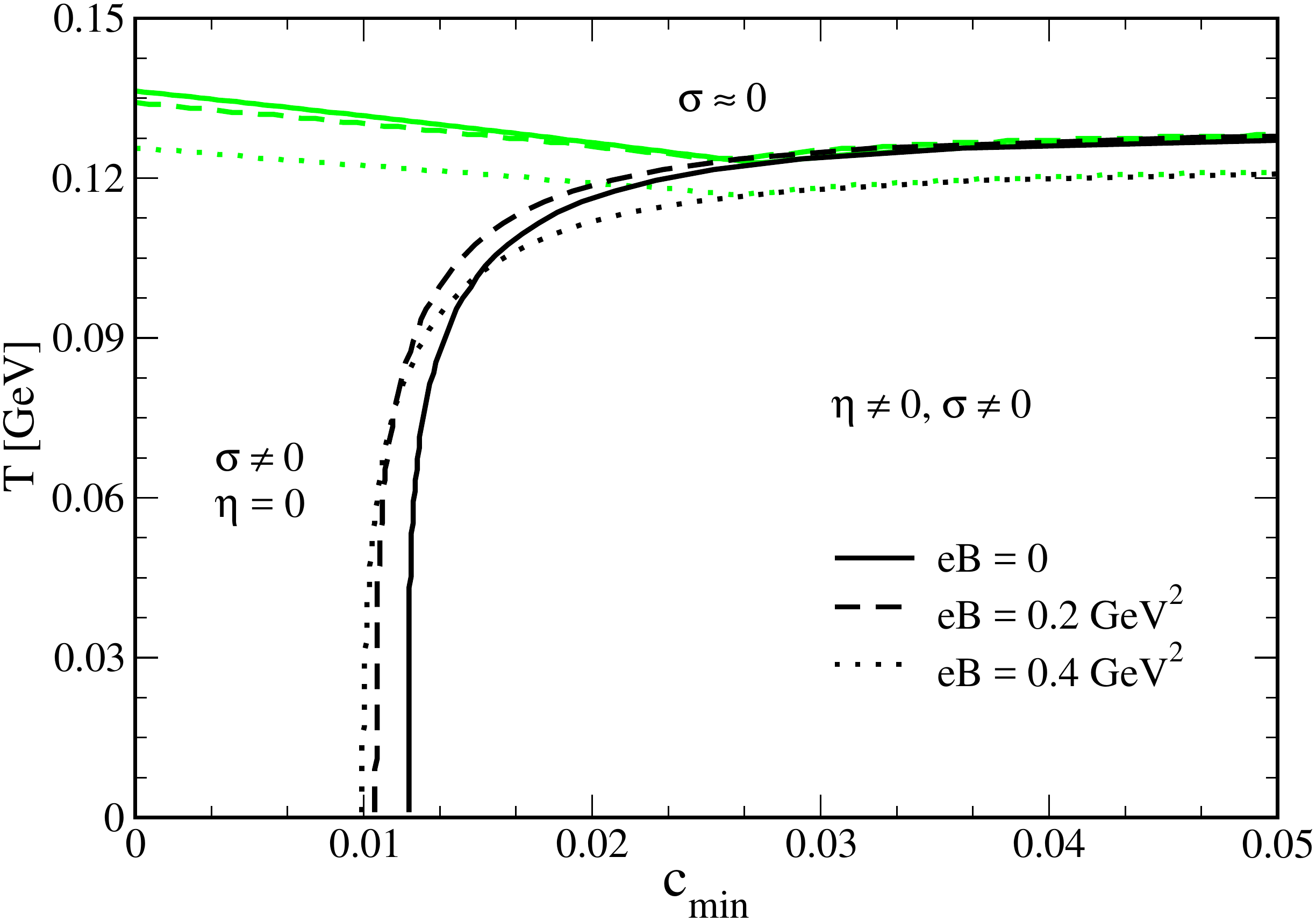}}
\caption{The $T - c_{\rm min}$ phase diagram, shown for three different values of external magnetic fields with the different parametrizations considered in this work. {}For these plots, we have considered the VEV of $a/f_a$ set as $\langle a\rangle/f_a = \pi$.}
\label{ccrit_vs_T}
\end{center}
\end{figure*}

{}Finally, for the choice of the parameter $c$ introduced in
subsection~\ref{subsec2a} that controls the strength of axion
interaction, there have been many discussions in the
past. Comparing with the three-flavor NJL model
parameters~\cite{Frank:2003ve} one gets the value of $c$ to be around
$0.2$. On the other hand, successful models like the instanton liquid
model, suggests $c\sim 1$, thereby indicating $\mathcal{L}_a$ as the
dominant part of the Lagrangian density. In the present work most of
our results consider $c=0.2$ following the
Refs.~\cite{Boomsma:2009eh,Lu:2018ukl}, though we have also shown a case
comparison for the topological susceptibility with two different
values of $c$, i.e., $c=0.2$ and $c=0.8$, which can be considered as
representative cases when comparing the results. 
{}Furthermore, choosing the vacuum expectation value (VEV) 
of the scaled axion field $a/f_a = \pi$ as an illustration, we have done a 
systematic analysis of the minimum value of $c$ (which we call as $c_{\rm min}$), 
i.e., the value of $c$ after which the condensates $\sigma$ and $\eta$ 
becomes nonvanishing, hence signalling the chiral and spontaneous $CP$ symmetry violating phases respectively. In {}Fig.~\ref{ccrit_vs_T}, 
we have shown the $T - c_{\rm min}$ phase diagram for different values 
of external magnetic field and with different parametrizations. From this phase diagram we can clearly identify the three different phases appearing in the $T - c_{\rm min}$ plane due to second order $CP$ phase transition and the chiral crossover. So the bottom right part of Fig.~\ref{ccrit_vs_T}, i.e., $\sigma\neq0,~\eta\neq0$ represents both the chiral and the spontaneous $CP$ symmetry violating phase, the bottom left part i.e., $\sigma\neq0,~\eta=0$ stands for the phase where chiral symmetry is still violated but spontaneous $CP$ symmetry is restored and finally the top part i.e., $\sigma\approx 0$ shows the almost chiral symmetric phase (chiral symmetry is not fully restored due to the nonvanishing quark masses). Figure~\ref{ccrit_vs_T} also shows that for each case corresponding to different values of $eB$ and parametrization, after certain values of higher temperatures (dependent on $eB$ and parametrization) the spontaneous $CP$ violating phase vanishes. These behaviors will be more transparent through the discussions of the next section. The results shown in {}Fig.~\ref{ccrit_vs_T} are also 
consistent with the ones shown for example, in Ref~\cite{Boomsma:2009eh}, 
for the values of current quark masses considered in our present work 
(i.e., given by Sets I, II and III). {}Finally all of the cases shown in  {}Fig.~\ref{ccrit_vs_T}
suggest  that the chosen value of $c=0.2$ is well within the viable regime. 

\section{Results and discussions}
\label{sec3}

In this section we will present our results for the different
quantities associated with the thermodynamics of the QCD axion in a
hot and magnetized medium within the NJL model. {}Firstly, we will
revisit the case with vanishing magnetic field~\cite{Lu:2018ukl} and
discuss some relevant points related to these results. Next, we will
move into the scenario with an external anisotropic magnetic
field. Then, we will deal with the two different procedures, both
related to the NJL model. {}First we will be working with a fixed coupling
constant $G_s$, which accounts for only MC. Then, finally, we will
also incorporate the IMC effect in our results using the effective $B$ and $T$
dependent 
coupling constant $G_s(eB,T)$ defined in the previous section, and
discuss about the effects of both MC and IMC on the axionic QCD
system.  

\subsection{The $eB = 0$ case}

The case of a vanishing magnetic field has recently been studied with
a different parametrization in Ref.~\cite{Lu:2018ukl}. In the current
study we will present some different aspects along with some of the
quantities already evaluated in Ref.~\cite{Lu:2018ukl}, for the
purposes of comparing between different parametrizations. This will be
particularly useful as a way of quantifying the differences that
different  parametrizations can make on the results.

\begin{center}
\begin{figure}[!htb]
\subfigure[]{\includegraphics[width=8cm]{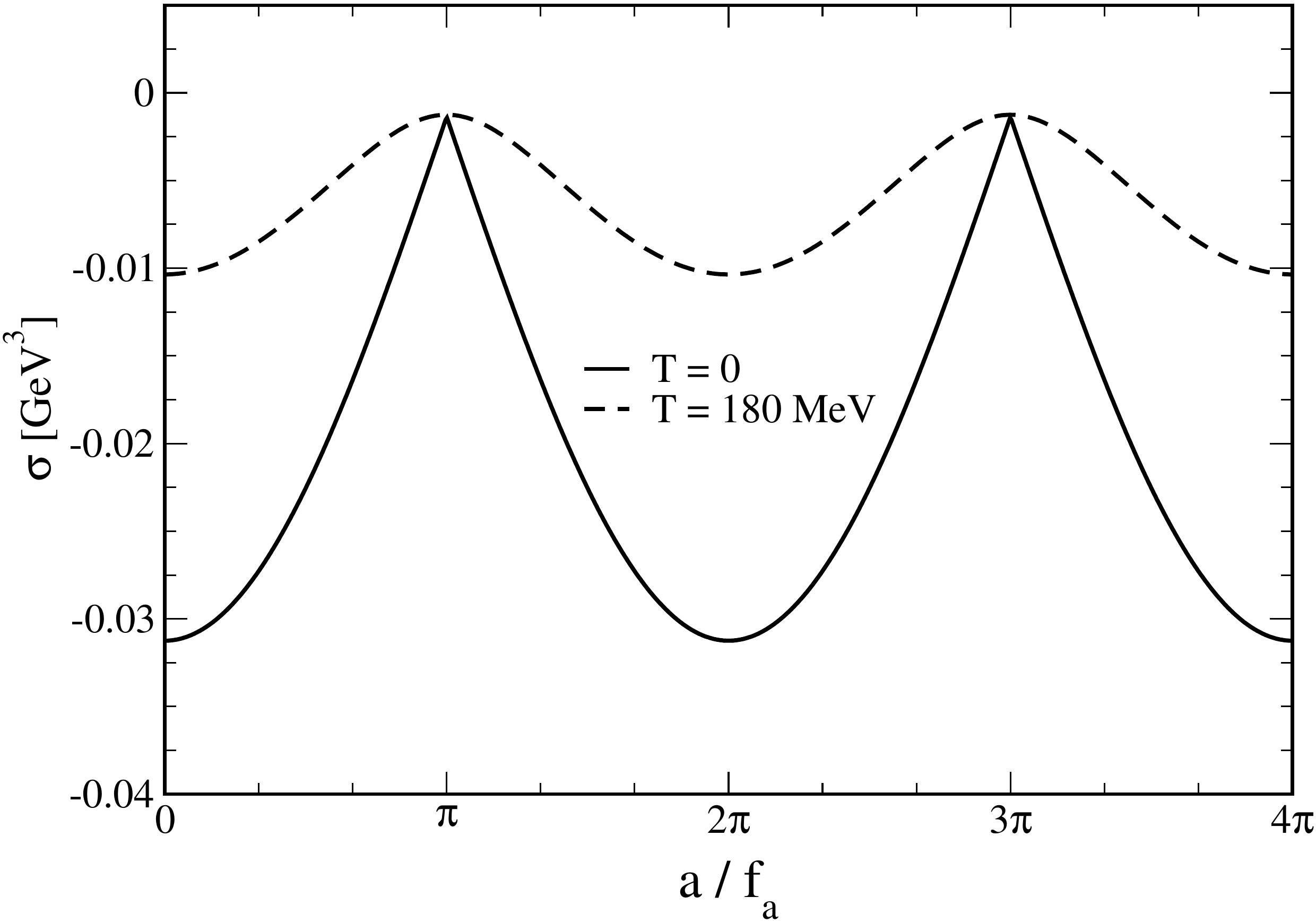}}
\subfigure[]{\includegraphics[width=8cm]{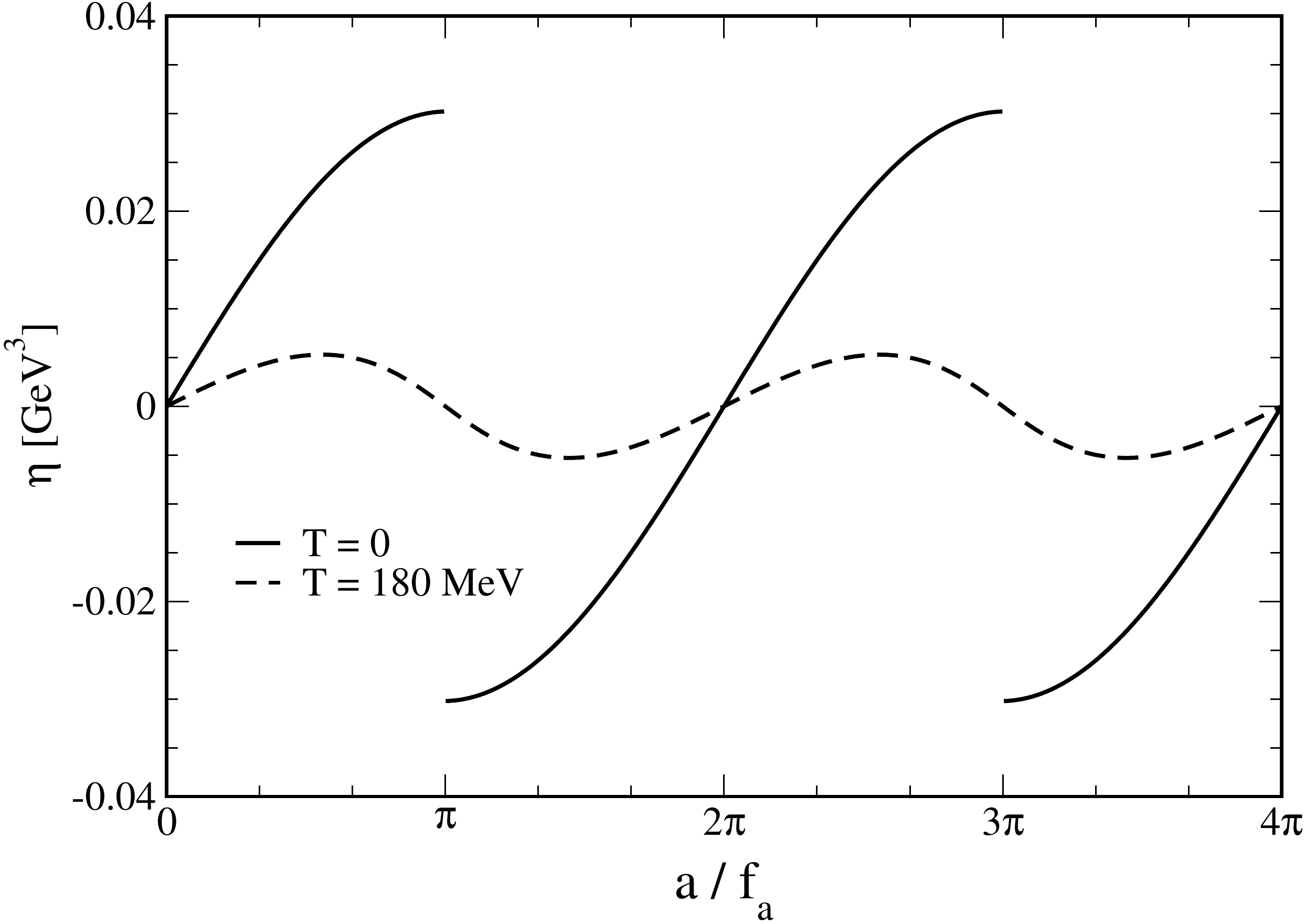}}
\caption{(a) Variation of $\sigma$ with respect to $a/f_a$ for two
  different values of $T$  and vanishing magnetic field. (b) Same
  variation done for $\eta$.  The curves are obtained using the 
  parameters from Set I shown in Table~\ref{parameter_sets}.}
\label{mf_vs_abyfa}
\end{figure}
\end{center}

In {}Fig.~\ref{mf_vs_abyfa} we have shown the variation of the
physical condensates $\sigma$ and $\eta$ with the axion field $a$
scaled with $f_a$ for both $T=0$ and $T=180$ MeV. The sinusoidal
behavior can be attributed to the $\sin(a/f_a)$ and $\cos(a/f_a)$
functions present in the effective mass $M$, appearing explicitly in
the defined quantities in Eqs.~(\ref{eff_mass})-(\ref{alpha_beta}). It
is evident from {}Fig.~\ref{mf_vs_abyfa} that the behaviors of
$\sigma$ and $\eta$ differ whether the results are obtained for $T=0$ 
or for $T\neq 0$. As one can note from {}Fig.~\ref{mf_vs_abyfa}(b), at $T=0$ 
$\eta$ displays discontinuities for $a/f_a = (2j+1)\pi$, for
$j=0,\,1,\,2, \ldots$, whereas for $a/f_a = 2j\pi$ it vanishes. 
This result agrees with the Dashen's phenomena~\cite{Dashen:1970et}, 
which predicts the existence of two degenerate vacua at $T=0$ and $a/f_a = (2j+1)\pi$ 
due to spontaneous $CP$ symmetry violation\footnote{Vafa-Witten theorem~\cite{Vafa:1984xg} 
dictates that for $a/f_a = 2j\pi$ the Lagrangian density is both explicitly and 
spontaneously $CP$ conserving. But for $a/f_a = (2j+1)\pi$, though the Lagrangian density
is explicitly $CP$ symmetric, no restrictions are imposed on the spontaneous 
$CP$ symmetry breaking.}.
Different signs of the two vacua indicates that they differ by a $CP$ transformation 
between them. But Dashen's phenomena starts to break down with the increase in 
temperature. After a certain critical temperature $T_c$, the spontaneous $CP$ symmetry 
is restored and from the $CP$ violating phase ($\eta\neq 0, ~\sigma\neq 0$), it returns to 
the phase of the ordinary chiral condensate ($\eta = 0, ~\sigma\neq 0$). This value of $T_c$ depends on the external magnetic field and parametrization. In fact, it can be realized from the phase diagram showed in Fig.~\ref{ccrit_vs_T} also, e.g., the solid curve in {}Fig.~\ref{ccrit_vs_T}(a) confirms the fact that for $eB=0$ and using the parameters from the Set I case, the spontaneous $CP$ violating phase has already been disappeared before $T=180$ MeV. This is evident from {}Fig.~\ref{mf_vs_abyfa}(b), looking at the dashed curve at $T=180$ MeV, where now we have only one vacuum. 
On the other hand $\sigma$ attains its minimum value for $a/f_a = 2j\pi$ and reaches the 
maximum value for $a/f_a = (2j+1)\pi$ for both $T=0$ and $T\neq 0$. 

\begin{center}
\begin{figure}[!htb]
\subfigure[]{\includegraphics[width=8cm]{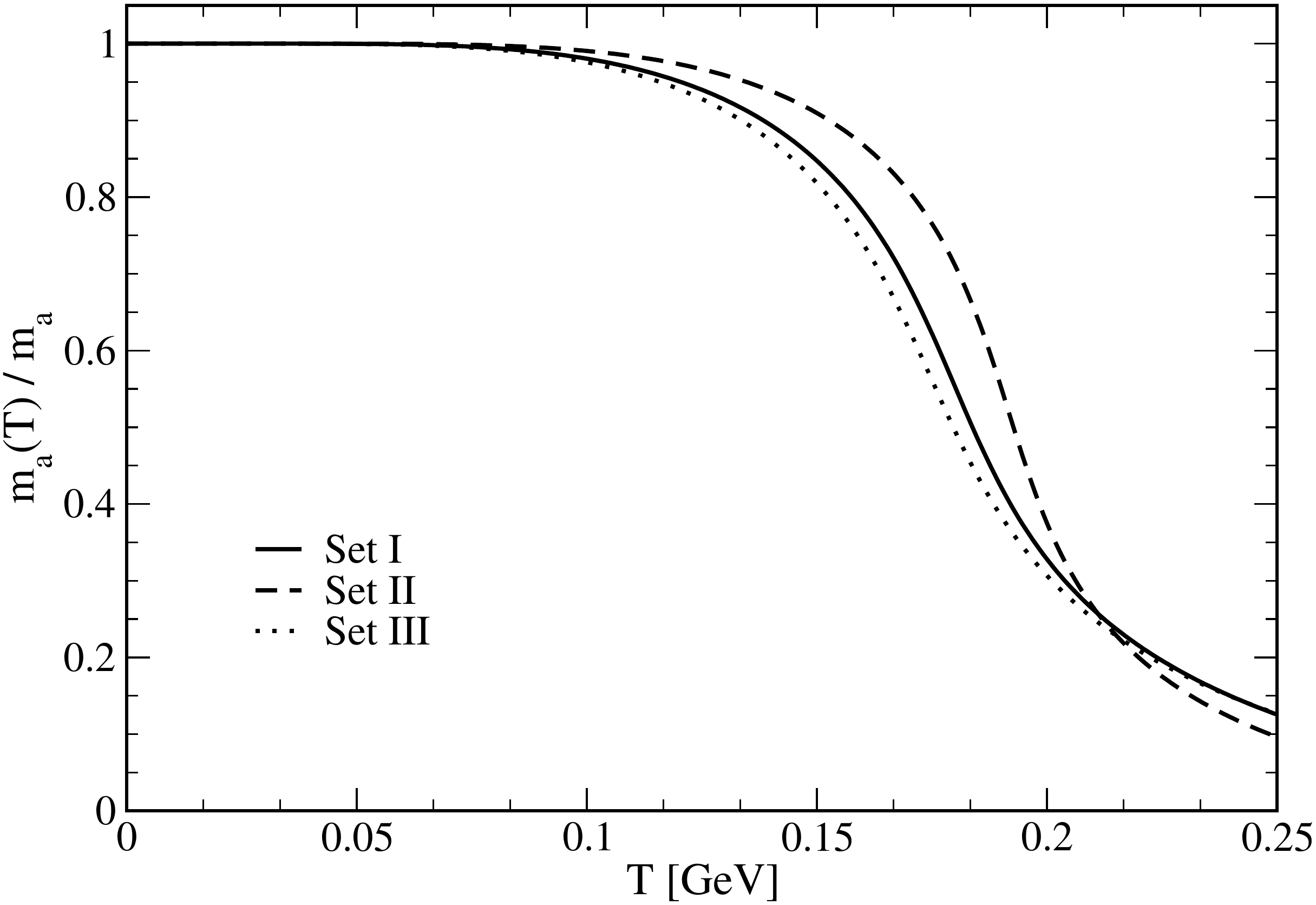}}
\subfigure[]{\includegraphics[width=8cm]{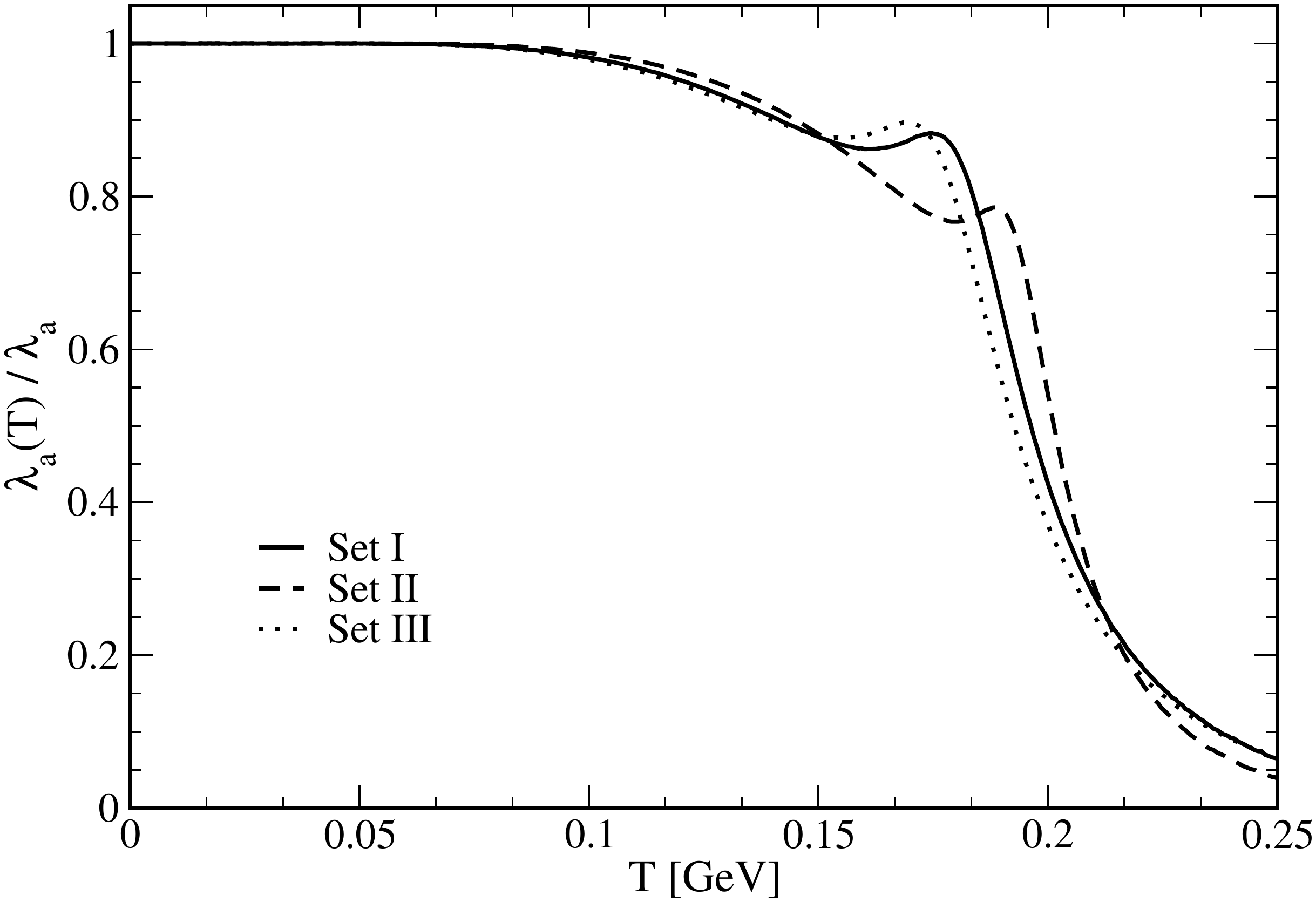}}
\caption{(a) Comparison between the variation of the axion mass ratio
  $m_a(T)/m_a$ with respect to $T$ for vanishing magnetic field with
  three different sets of parametrizations (see
  Table~\ref{parameter_sets}). (b) Same comparison done for the axion
  self-coupling ratio $\lambda_a(T)/\lambda_a$.}
\label{ma_lambdaa_pc}
\end{figure}
\end{center}

In {}Fig.~\ref{ma_lambdaa_pc} we have shown the comparison between
three different sets of parametrizations listed in
Table~\ref{parameter_sets} for the variations of the temperature
dependent axion mass $m_a(T)$ and the temperature dependent axion
self-coupling $\lambda_a(T)$, scaled with their respective zero
temperature values. One notices that after the value of $T\simeq 0.1$
GeV, the different parametrizations start to deviate from each
other. {}For both the case of the axion mass and the axion
self-coupling, a relatively rapid decrease is noticed around the
chiral (pseudo critical) transition temperature $T_c$, while for the self-coupling a
kink-like feature appears in this region.

\begin{center}
\begin{figure}[!htb]
\subfigure[]{\includegraphics[width=8cm]{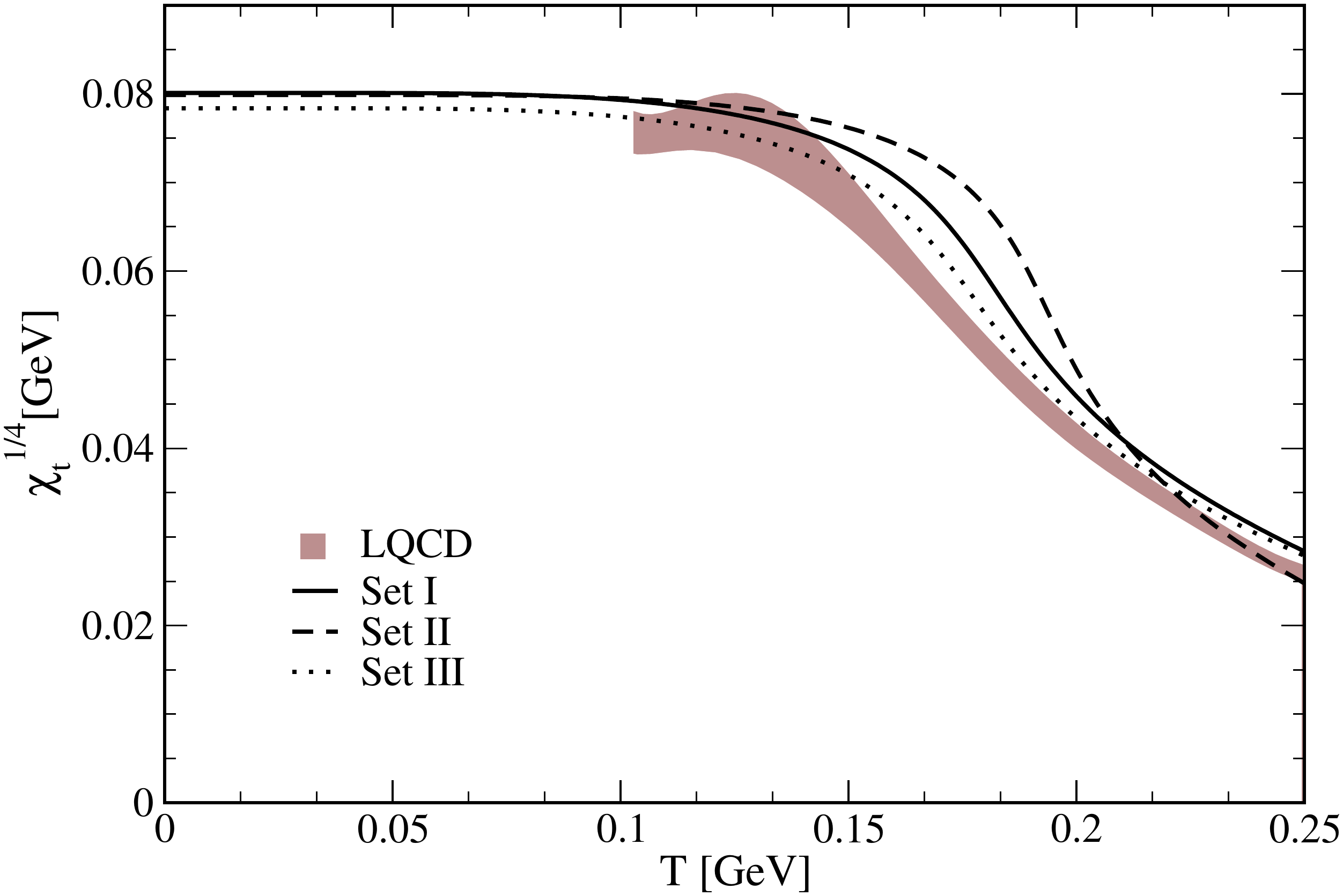}}
\subfigure[]{\includegraphics[width=8cm]{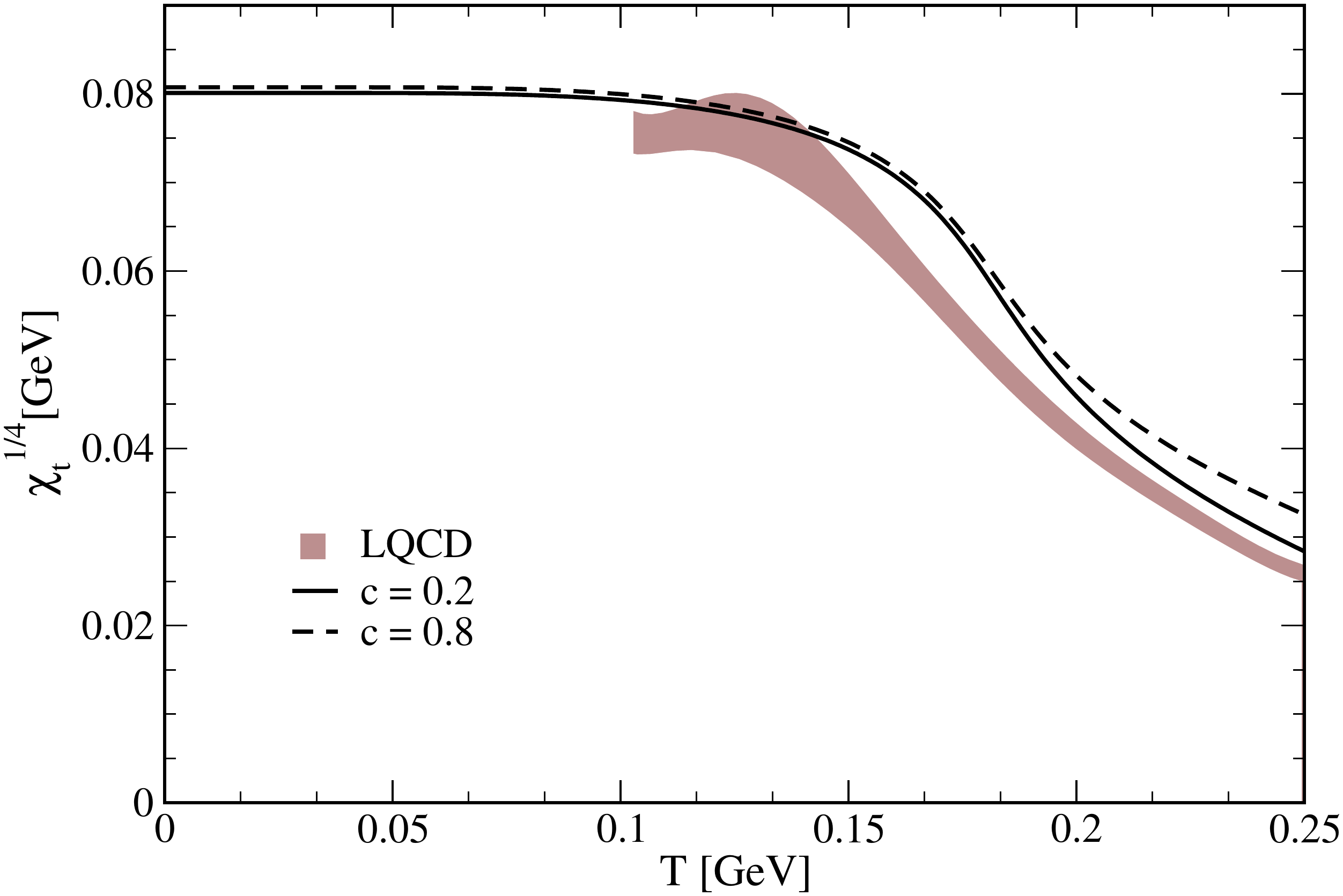}}
\caption{(a) Comparison between the variation of the topological
  susceptibility $\chi_{	\rm{top}}$ with respect to $T$ and for
  vanishing magnetic field with three different parametrizations (see
  Table~\ref{parameter_sets}) along with the lattice QCD results
  from~\cite{Borsanyi:2016ksw}. (b) Comparison of $\chi_{
    \rm{top}}$ with respect to $T$ with two different values for the
  constant $c$, i.e., two different strengths of the axion
  interactions along with the lattice QCD results. In this case it is
  used the parameters from Set I shown in Table~\ref{parameter_sets}.}
\label{topsus_pc}
\end{figure}
\end{center}

In {}Fig.~\ref{topsus_pc} we show the results for the topological
susceptibility $\chi_{	\rm{top}}(T)$, for which we also have
available the lattice QCD results~\cite{Borsanyi:2016ksw} at zero
magnetic field. We have displayed the variation of $\chi_{
  \rm{top}}$ with respect to the temperature for the three different
parametrizations sets previously mentioned and compared them both with
the available lattice QCD results. As one can see from
{}Fig.~\ref{topsus_pc}(a), the parameter sets I and III given in the
Table~\ref{parameter_sets} look more compatible with the lattice
data. In this case,  we can also see that $\chi_{\rm{top}}(T)$ has
decreased substantially around $T_c$. This is a behavior matched by
the corresponding lattice QCD data. In {}Fig.~\ref{topsus_pc}(b), we
have plotted $\chi_{	\rm{top}}(T)$ in the case of choosing two
different values of the constant $c$, corresponding to two different
strengths of the axion interaction. When the strength of the axion
interaction is weaker, like in the case where $c=0.2$, the topological
susceptibility is dragged down a little bit more at higher $T$ than in
the higher strength case of $c=0.8$. 

\subsection{The case of finite magnetic field, $eB \neq 0$, but fixed
  coupling constant $G_s$}

In this subsection we work with the same parametrization used for the
$eB=0$ case, i.e., with a fixed coupling constant $G_s=5.022$
GeV$^{-2}$. As discussed earlier, with the fixed coupling constant we
only consider the effects of magnetic catalysis in this case. Below we
will also consider the more physical scenario, which accounts for the
inverse magnetic catalysis effect that appears around the pseudo critical 
temperature.

\begin{center}
\begin{figure}[!htb]
\subfigure[Results for
  $T=0$.]{\includegraphics[width=8cm]{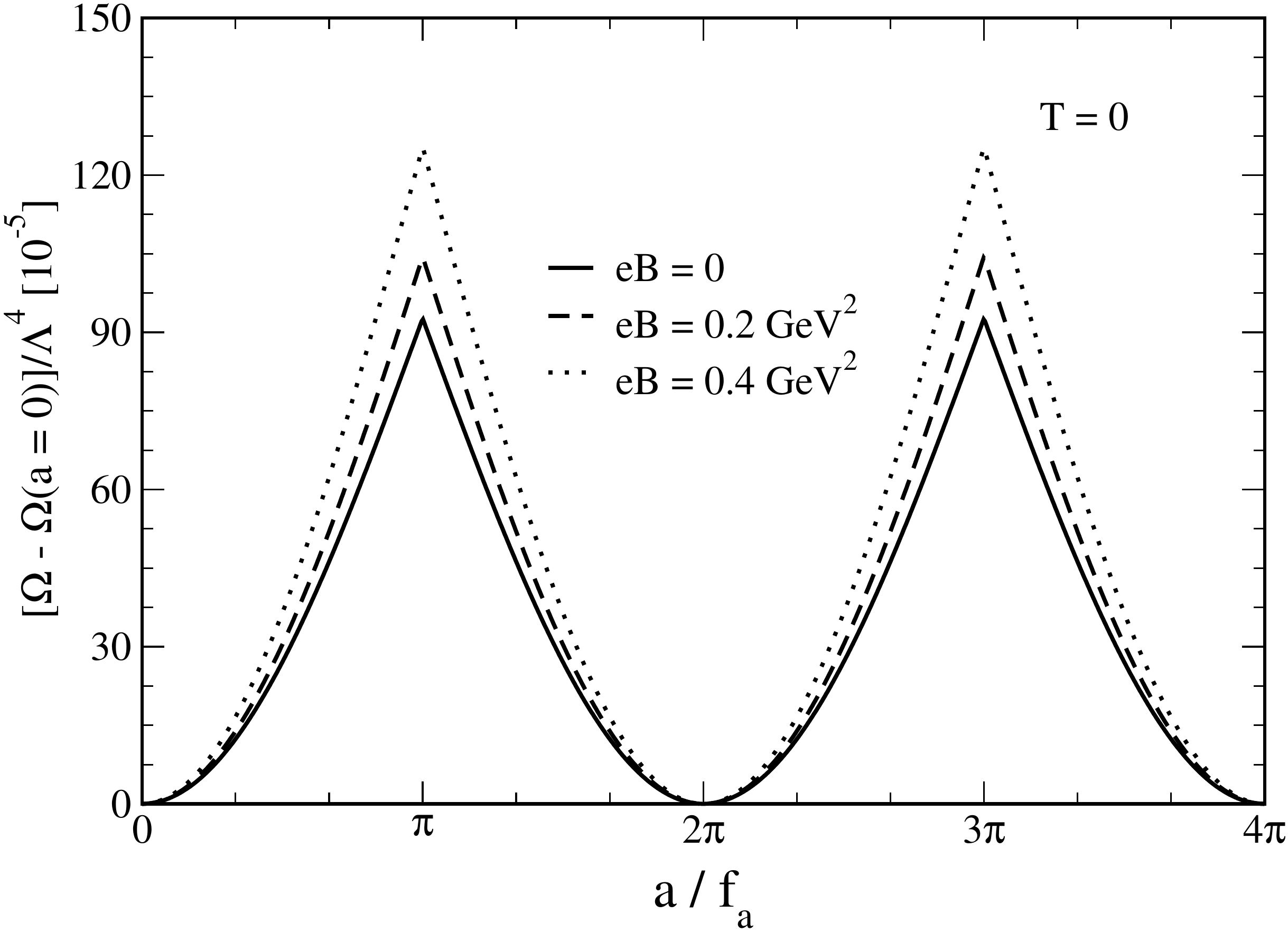}}
\subfigure[Results for $T=180$
  MeV.]{\includegraphics[width=8cm]{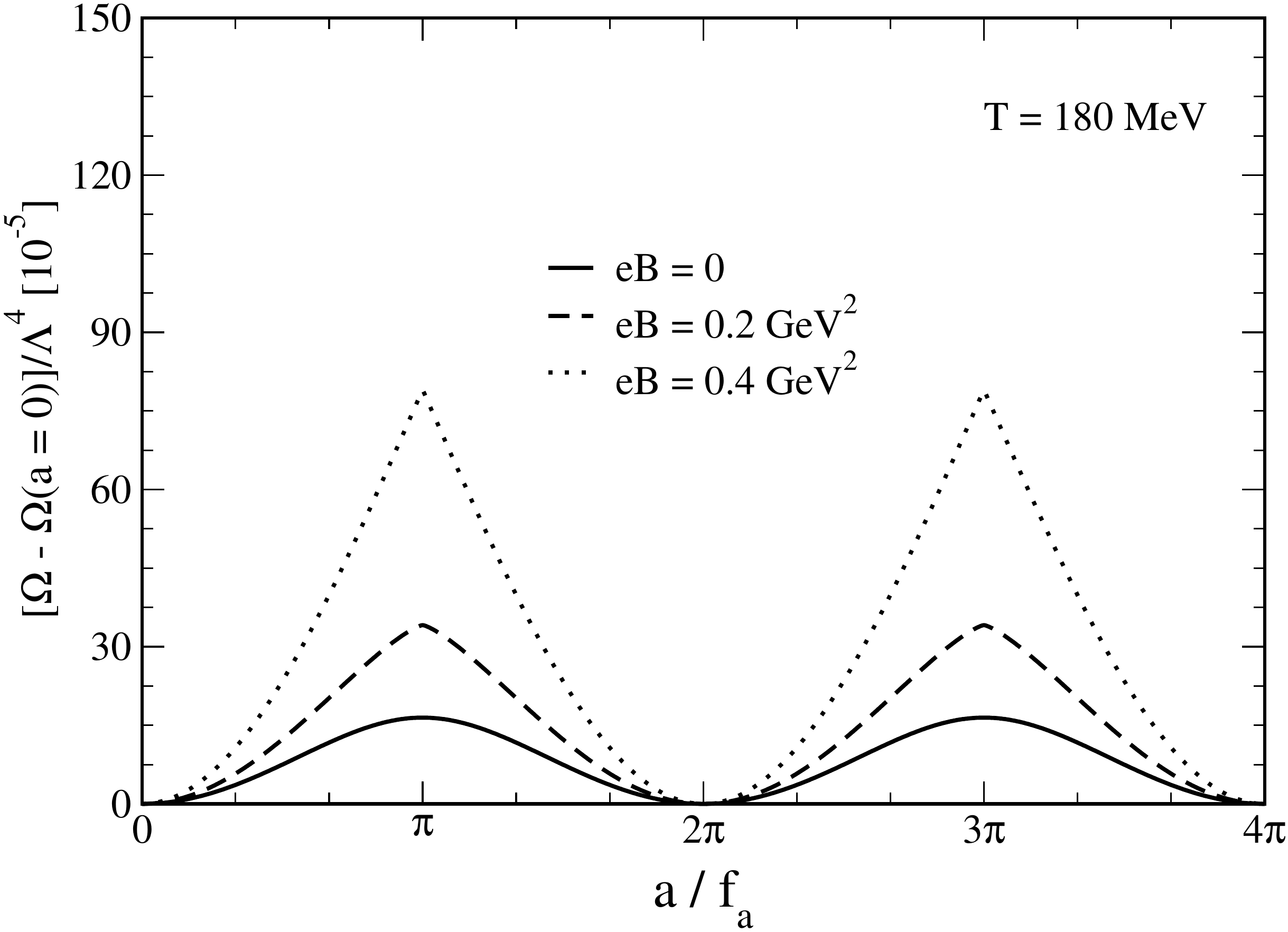}}
\caption{Variation of the scaled axion contribution of the effective
  thermodynamic potential $\Omega-\Omega(a=0)$ with the scaled axion
  field $a/f_a$ for three different values of external magnetic
  field. The curves are obtained using the parameters from Set I shown
  in Table~\ref{parameter_sets} for the magnetized medium and for two
  different values of temperature, i.e., for $T=0$ (a) and for $T=180$
  MeV (b).}
\label{Omega_fixedG}
\end{figure}
\end{center}

In {}Fig.~\ref{Omega_fixedG}, the variation of the scaled axionic part
of the effective thermodynamic potential is shown with the scaled
axion background field $a/f_a$ for three different values of the
external magnetic field, namely, $eB=0, 0.2$ GeV$^2$ and $0.4$
GeV$^2$, respectively. The aforementioned variation is shown for two
different temperatures in the two panels shown in
{}Fig.~\ref{Omega_fixedG}. It is apparent from these results that for both $T=0$ and
$T=180$ MeV, with increasing magnetic field the value of the effective
thermodynamic potential gets increased, which effectively also follows
a valley-hill like structure. This behavior can be traced again due to
the presence of the terms $\sin(a/f_a)$ and $\cos(a/f_a)$ in the
effective mass $M$. It is also noticeable that with increasing
temperature, the thermal effects start to melt down the amplitude of
the axionic part of the thermodynamic potential, which is visible most
prominently when we compare the case of $eB=0$ in Fig.~\ref{Omega_fixedG} (a) 
and Fig.~\ref{Omega_fixedG} (b) .

\begin{center}
\begin{figure}[!htb]
\subfigure[Case for
  $a/f_a=0$.]{\includegraphics[width=8cm]{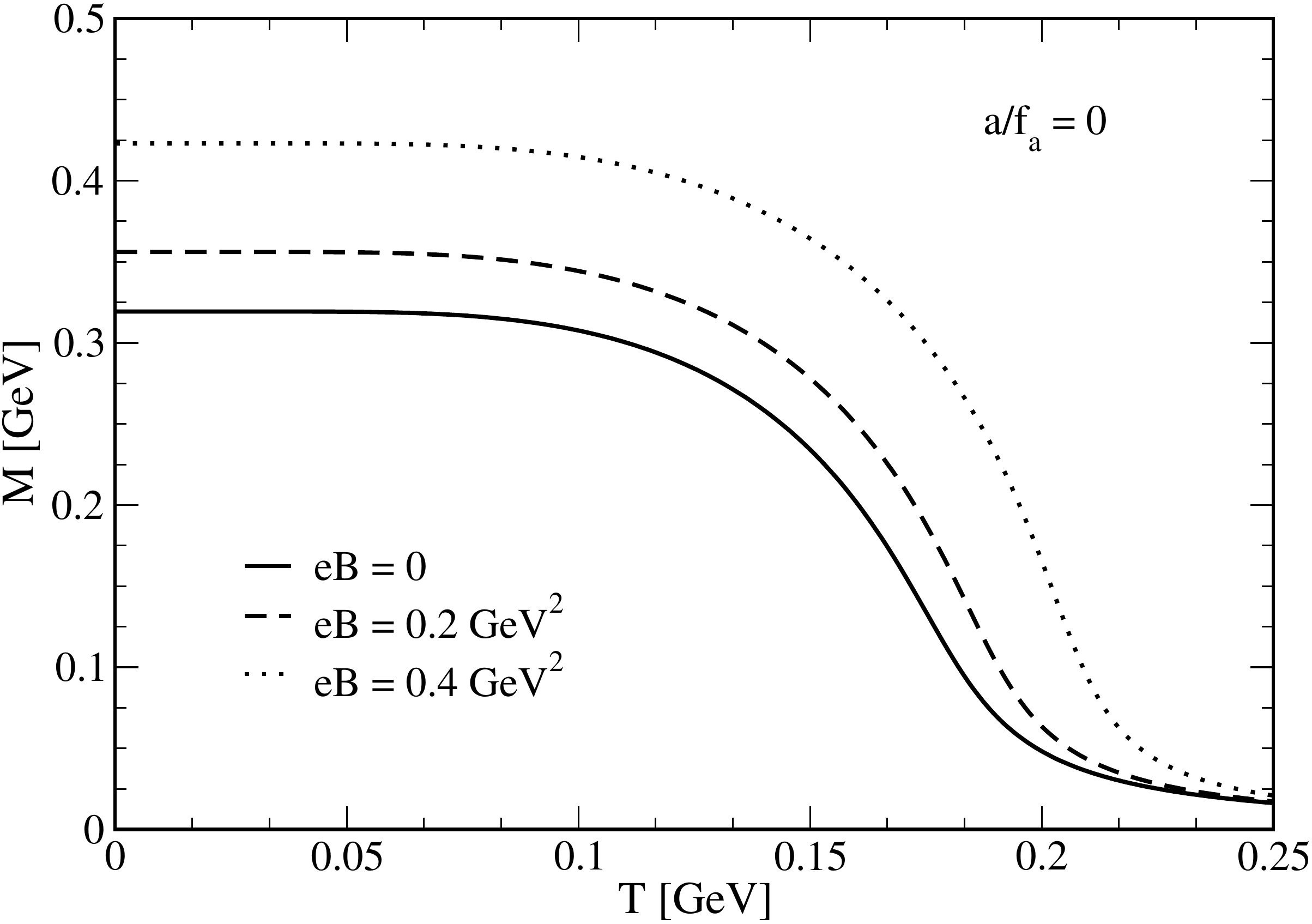}}
\subfigure[Case for for
  $a/f_a=2\pi/3$.]{\includegraphics[width=8cm]{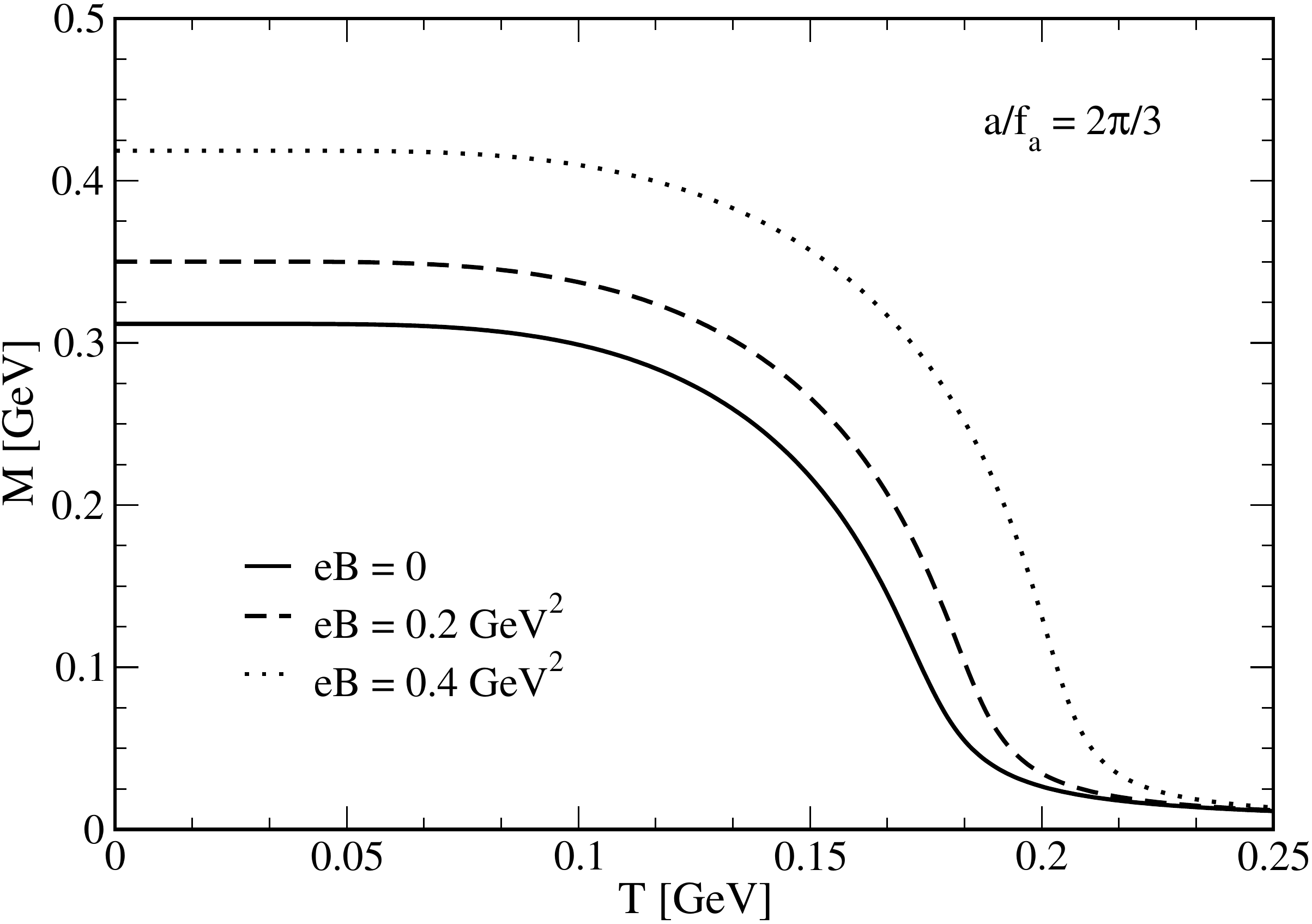}}
\caption{Variation of the constituent quark mass $M$ with $T$ for
  three different values of the magnetic field. The plots are done
  with the parameters from Set I given in the
  Table~\ref{parameter_sets} and for two different values of the
  scaled axion field, i.e., for $a/f_a=0$ and for $a/f_a=2\pi/3$.}
\label{eff_mass_fixedG}
\end{figure}
\end{center}

In {}Fig.~\ref{eff_mass_fixedG} we show the effects of the temperature
and magnetic field on the effective dynamical mass $M(T,B)$, which in
turn depends on the physical condensates $\sigma(T,B)$ and
$\eta(T,B)$. The effect of MC is clearly evident from the results
displayed in the figure: By increasing the magnetic field magnitude
$eB$, the effective mass increases throughout the temperature range
$T=0 \,- \,250$ MeV. The effect of the axion field is also noticeable
comparing the two plots in {}Fig.~\ref{eff_mass_fixedG} as we see that
the effective dynamical mass $M(T,B)$ gets slightly decreased with a
finite value of the scaled axion field. 

\begin{center}
\begin{figure}[!htb]
\subfigure[]{\includegraphics[width=8cm]{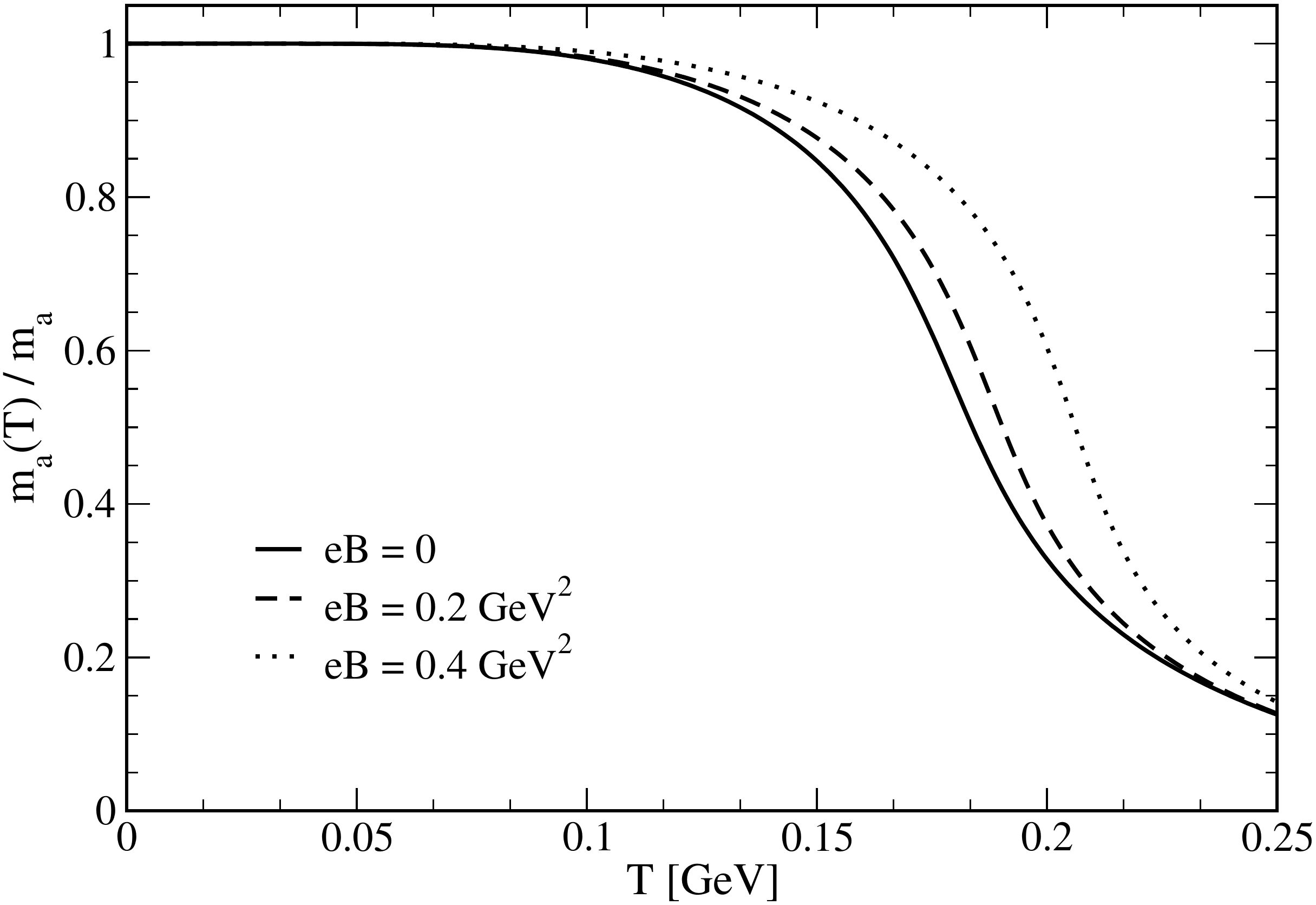}}
\subfigure[]{\includegraphics[width=8cm]{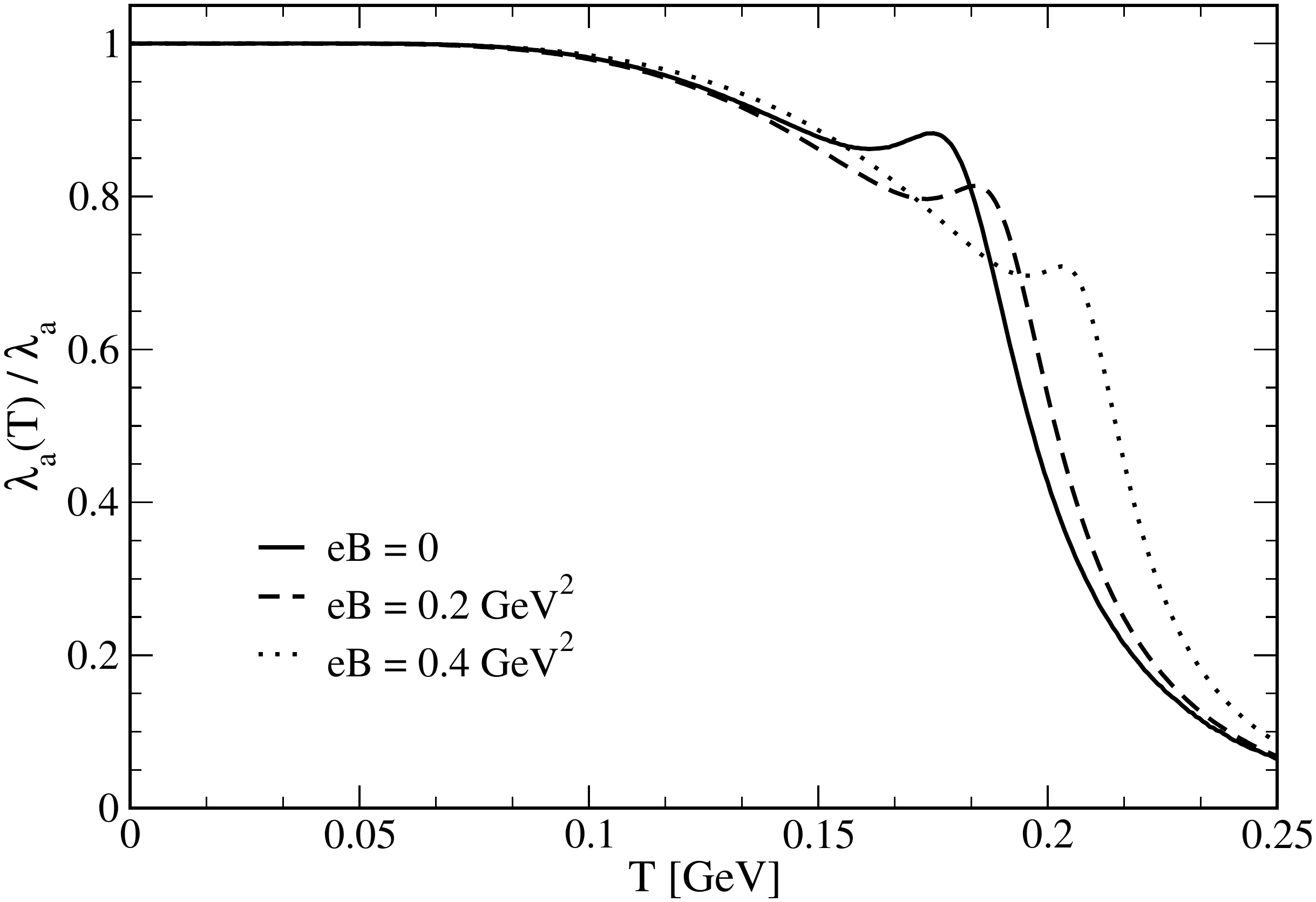}}
\subfigure[]{\includegraphics[width=8cm]{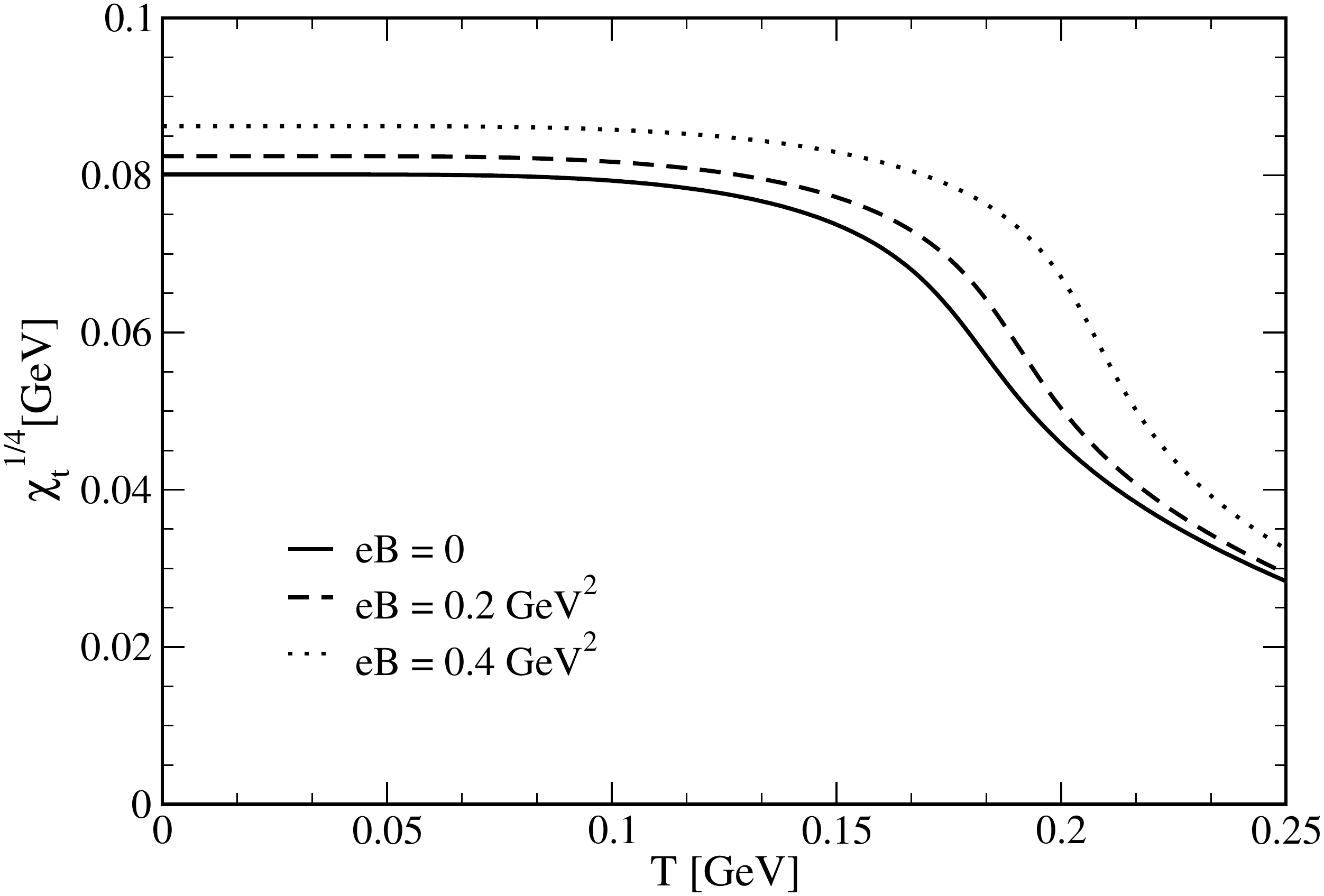}}
\caption{Panel (a): Variation of the axion mass ratio $m_a(T)/m_a$
  with respect to $T$ for three different values for the external
  magnetic field. Panel (b):  Same variation done for the axion
  self-coupling ratio $\lambda_a(T)/\lambda_a$. Panel (c): Likewise
  for the topological susceptibility $\chi_{
    \rm{top}}$. The plots are done with the parameters from Set I
  shown in the Table~\ref{parameter_sets}.}
\label{ma_lambda_fixedG}
\end{figure}
\end{center}

In {}Fig.~\ref{ma_lambda_fixedG} it is shown the temperature behavior
for the axion mass and for the self-coupling, both scaled with their
respective zero temperature values, for three different values for the
external magnetic field. {}For both the cases shown in the figure,
the magnetic catalysis
effects are visible around the inflection point close to $T_c$. We can
see the inflection points shifting towards higher values of $T$ for
higher values of $eB$, a clear sign of MC. Similar behavior of MC is
obtained in {}Fig.~\ref{ma_lambda_fixedG}(c), where it is shown the
topological susceptibility as a function of the temperature. The MC
effect is visible throughout the temperature range. We can notice that
the value of $\chi_{	\rm{top}}^{1/4}$ at zero temperature goes up
to $\sim 0.0865$ GeV for the highest value of $eB$ considered,
$eB=0.4$ GeV$^2 \sim 20 m_\pi^2$.

\subsection{The case of considering a temperature and magnetic
  field dependent coupling $G_s(eB,T)$ constrained by lattice QCD}

{}Finally, let us here explore the more physical scenario involving
the effect of inverse magnetic catalysis around the transition
temperature, originally predicted by lattice QCD for a hot and
magnetized medium. As mentioned previously, for the purposes of incorporating IMC
in the ambit of the NJL model, in this subsection we have used a
temperature and magnetic field dependent coupling constant $G_s(eB,T)$.

\begin{center}
\begin{figure}[!htb]
\subfigure[Case for
  $T=0$.]{\includegraphics[width=8cm]{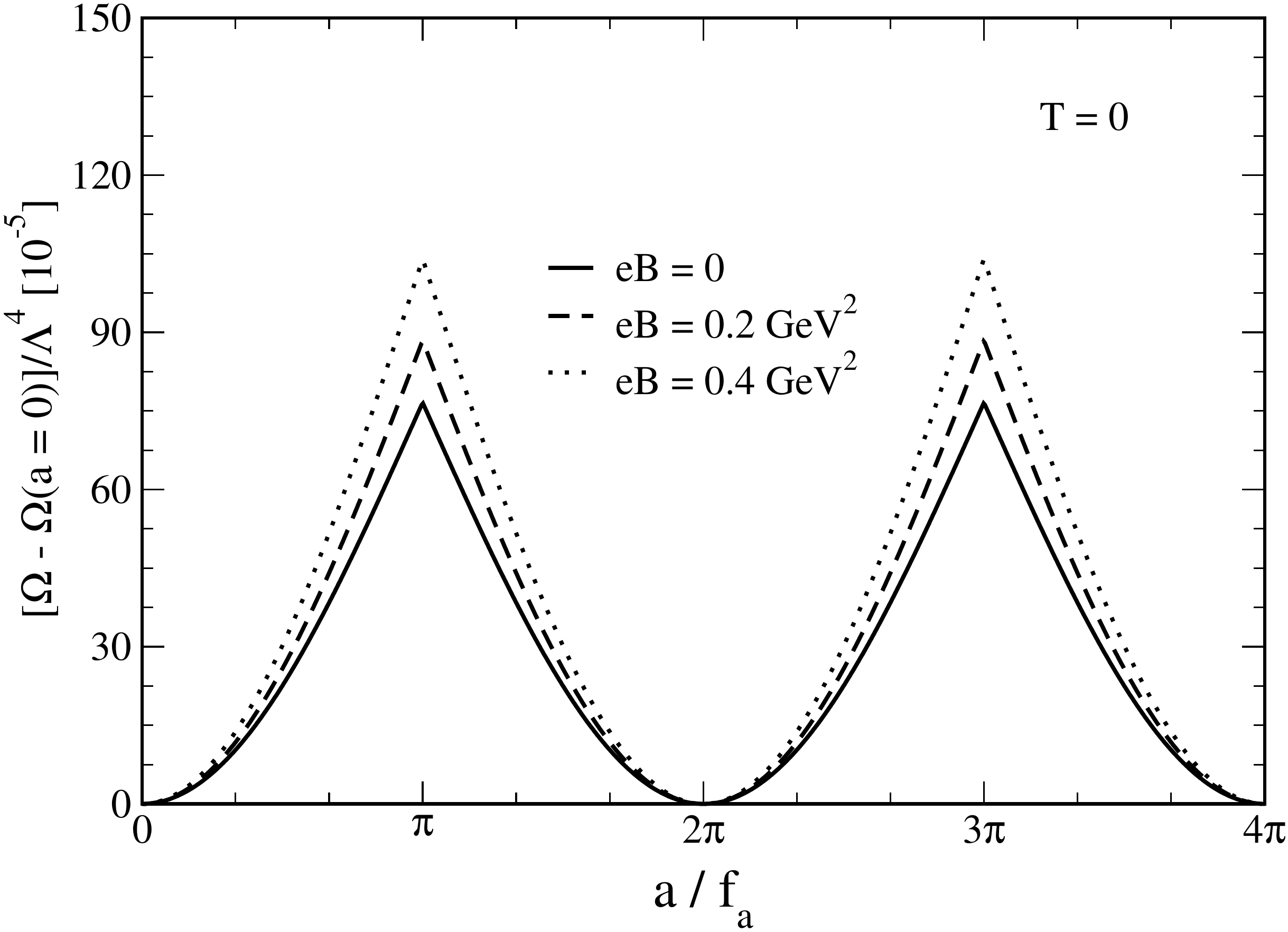}}
\subfigure[Case for $T=180$
  MeV.]{\includegraphics[width=8cm]{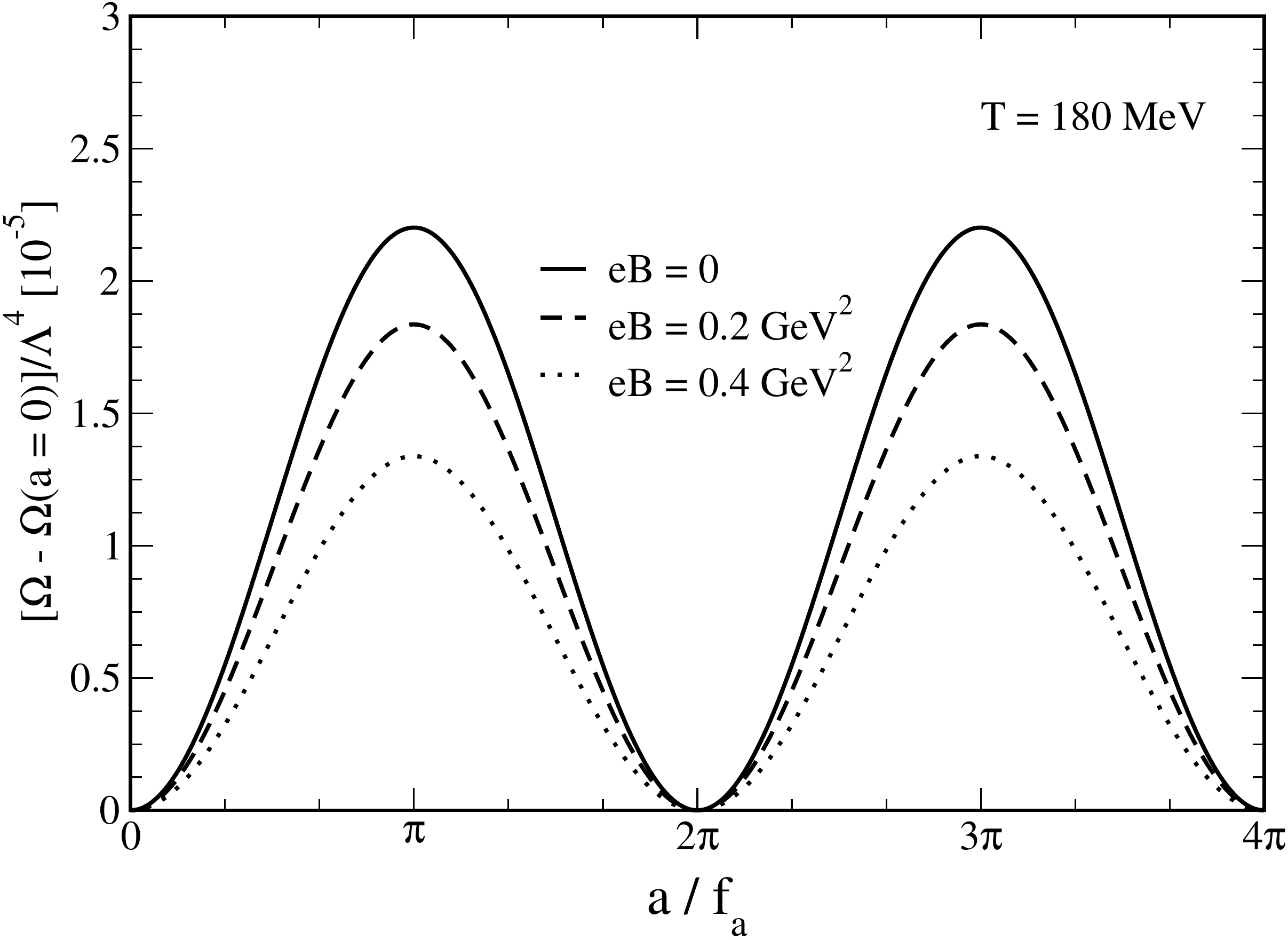}}
\caption{Same as in {}Fig.~\ref{Omega_fixedG}, but considering now the
  effect of a $eB$ and $T$ dependent coupling constant $G_s(eB,T)$ for
  two different temperatures: Panel (a) $T=0$ and panel
  (b) $T=180$ MeV.}
\label{Omega_varyG}
\end{figure}
\end{center}

In {}Fig.~\ref{Omega_varyG} we have again plotted the scaled axion
dependent part of the thermodynamic potential varying with respect to
the scaled axion field. This time, using the fully temperature and magnetic
field dependent coupling constant $G_s(eB,T)$, given by
Eq.~(\ref{imc_fit}). The main qualitative and quantitative difference in this case
with respect to that shown in the previous {}Fig.~\ref{Omega_fixedG} can
be noticed in {}Fig.~\ref{Omega_varyG}, where the effect of IMC is
evident throughout the whole range of the axion background
field. {}For the case of $T=180$ MeV, i.e., $T\sim T_c$, we can see
that with increasing magnetic field the amplitude of the thermodynamic
potential gets decreased. {}For this case, the overall value of the
scaled thermodynamic potential also becomes very low because of the
low value of $G_s(eB,T)$ at higher temperatures. The strong dependence
on the way the coupling constant is parametrized is evident comparing
{}Fig.~\ref{Omega_fixedG} with {}Fig.~\ref{Omega_varyG}.

\begin{center}
\begin{figure}[!htb]
\subfigure[Case for
  $a/f_a=0$.]{\includegraphics[width=8cm]{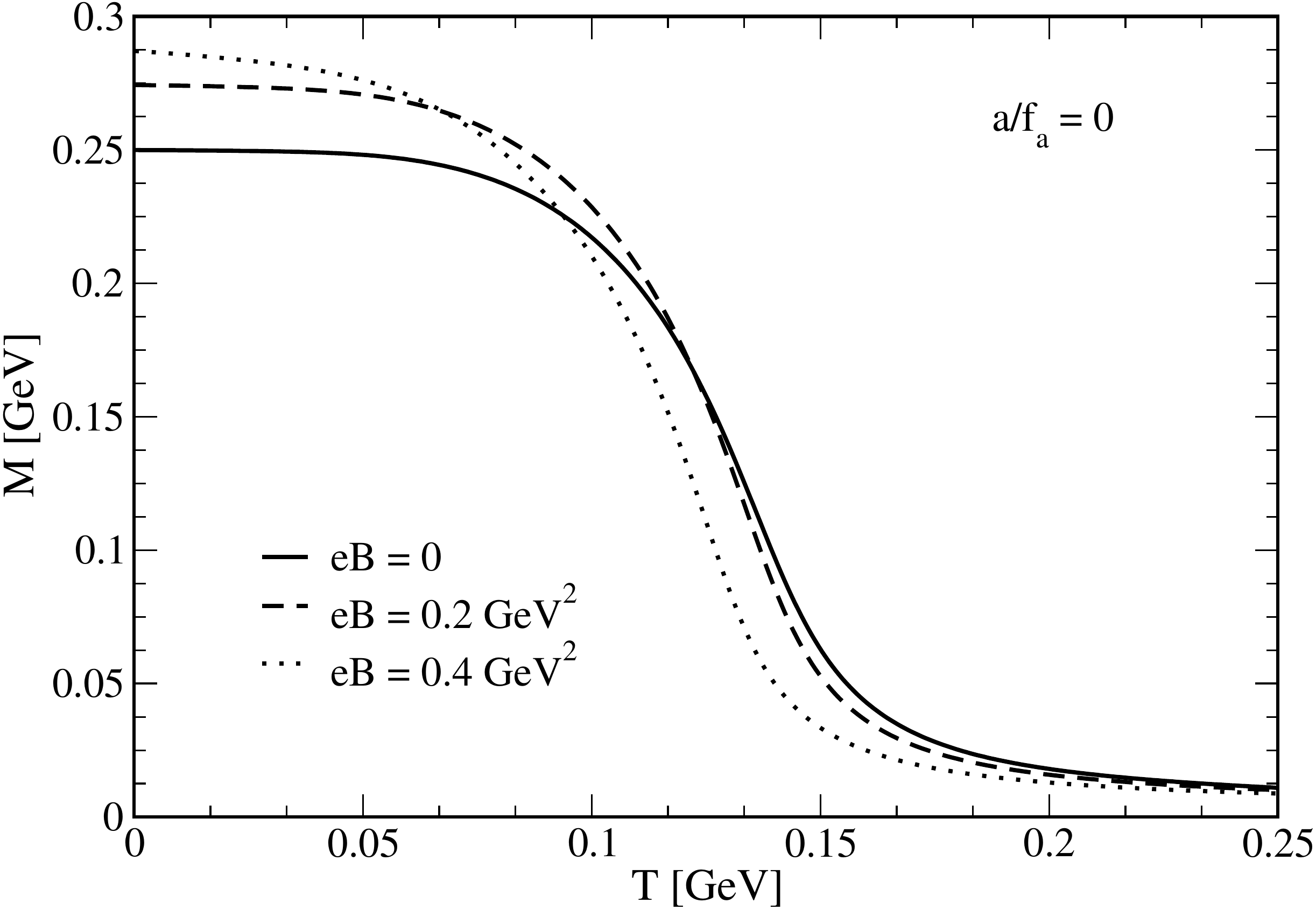}}
\subfigure[Case for $a/f_a=2\pi/3$.]{\includegraphics[width=8cm]{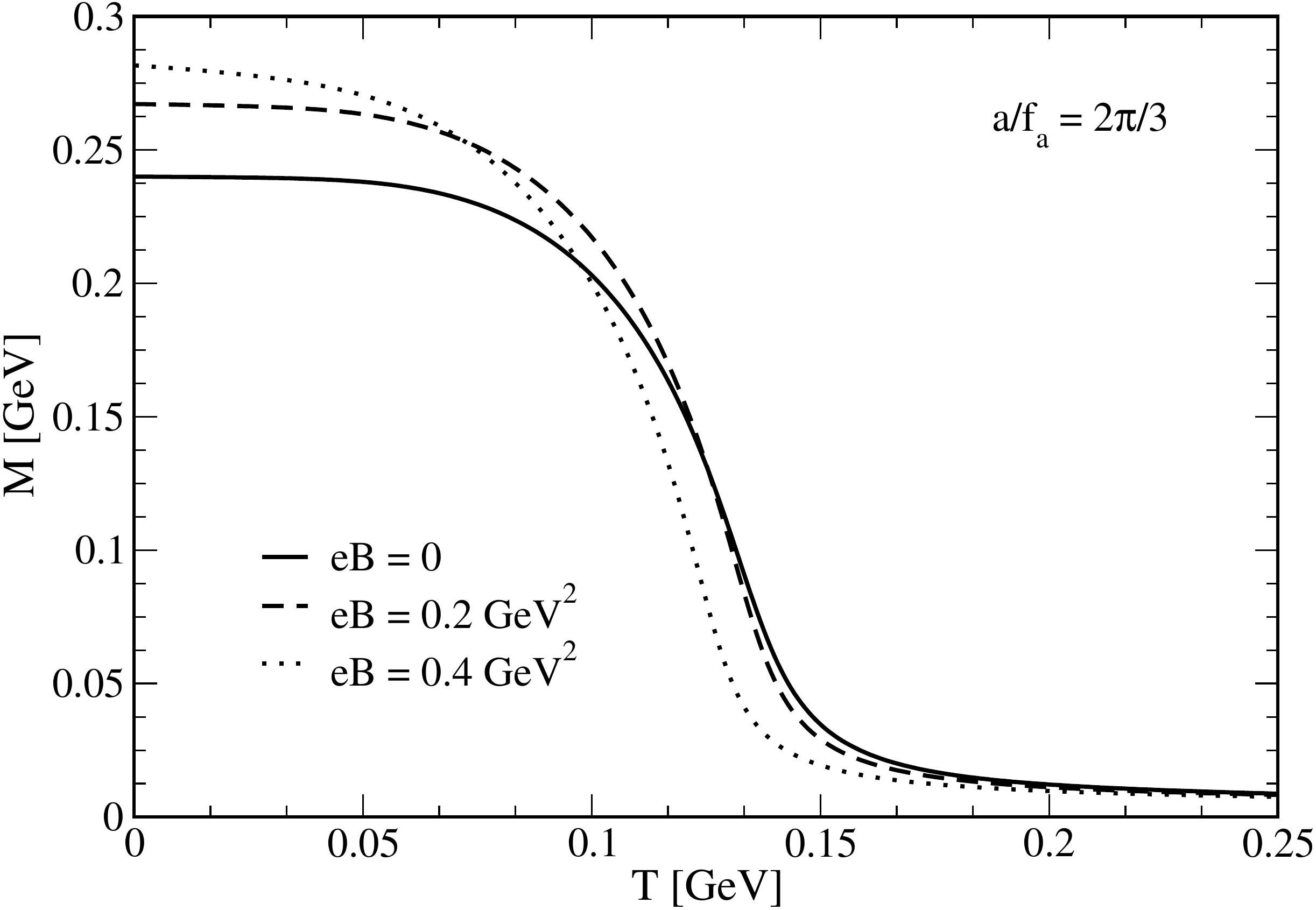}}
\caption{Same as shown in {}Fig.~\ref{eff_mass_fixedG}, but
  considering the $eB$ and $T$ dependent coupling constant $G_s(eB,T)$
  for two different values of the scaled axion field, i.e. $a/f_a=0$
  and $a/f_a=2\pi/3$.}
\label{eff_mass_varyG}
\end{figure}
\end{center}

In {}Fig.~\ref{eff_mass_varyG} we show the variation of the effective
constituent quark mass $M$ with respect to $T$ and using the explicitly
dependent on $B$ and $T$ coupling constant $G_s(eB,T)$. 
It clearly explores the full spectrum of
hot magnetized medium showing both the MC and IMC effects for both
vanishing and finite axion field. At zero to lower temperatures it
shows the catalysis effect as higher $eB$ values correspond to higher
effective mass. But at the higher temperatures, it goes through a
crossover and around $T_c$ the behavior is completely opposite,
emphasizing the occurrence of the inverse magnetic catalysis in that
regime.

\begin{center}
\begin{figure}[!htb]
\subfigure[]{\includegraphics[width=8cm]{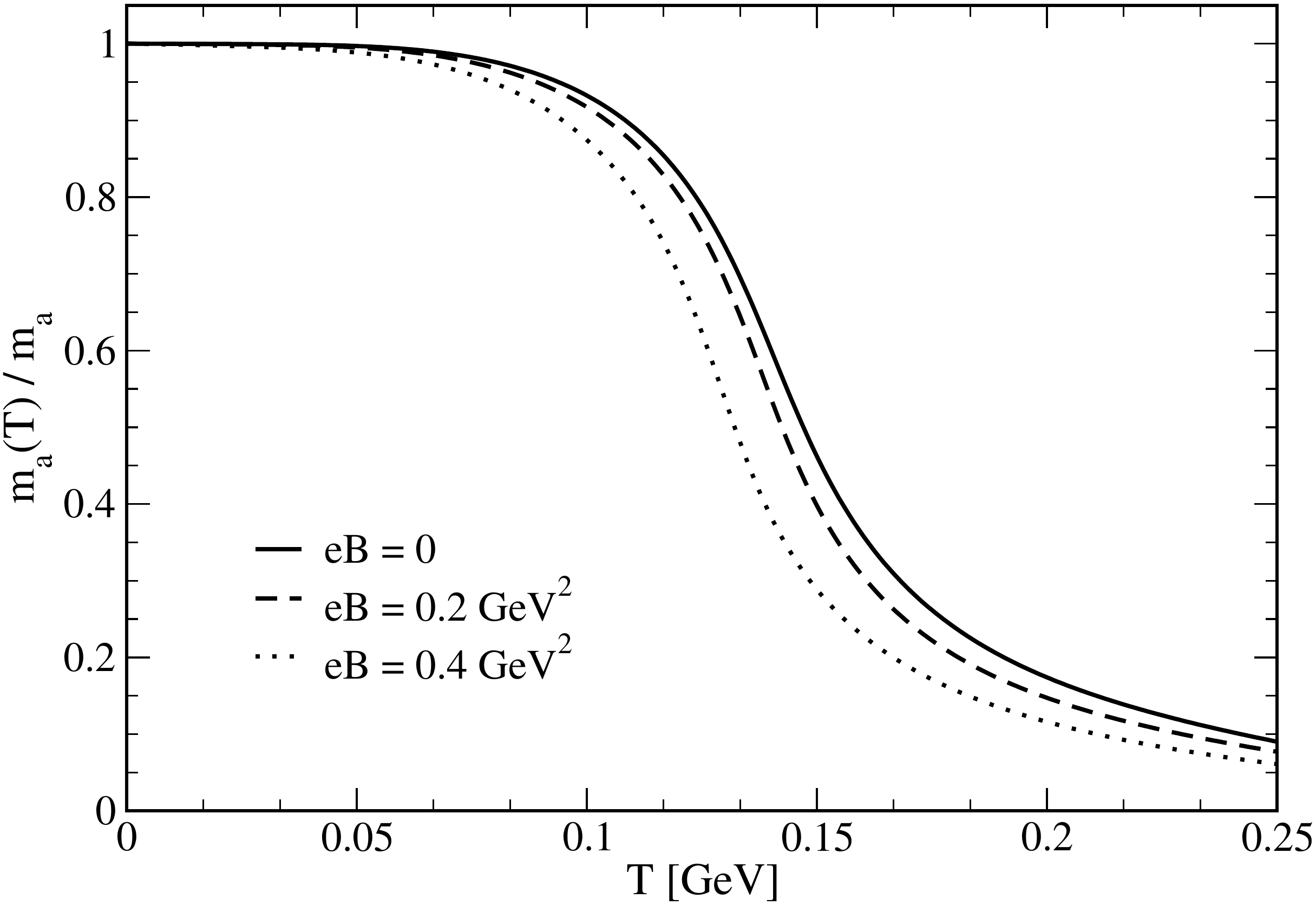}}
\subfigure[]{\includegraphics[width=8cm]{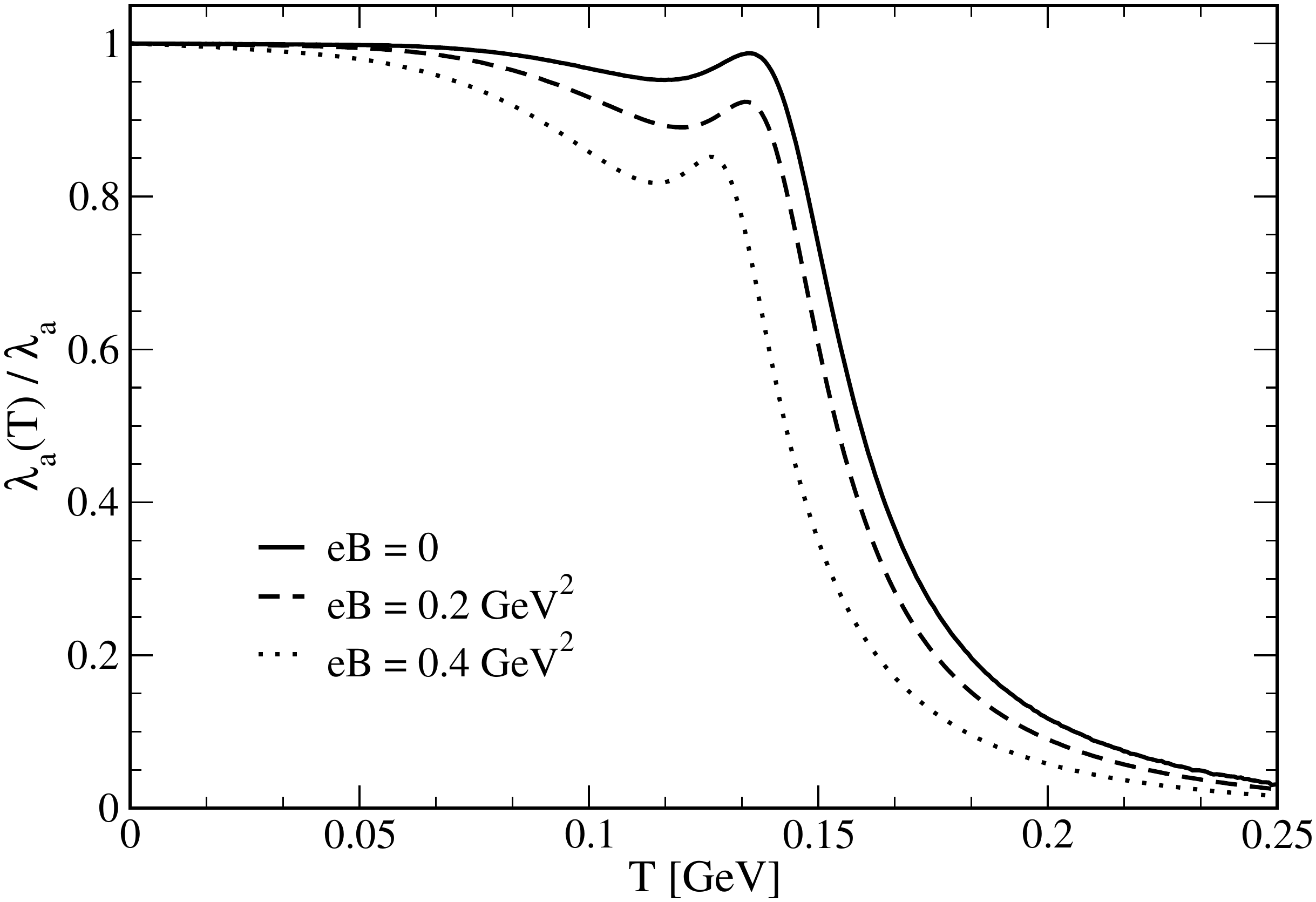}}
\subfigure[]{\includegraphics[width=8cm]{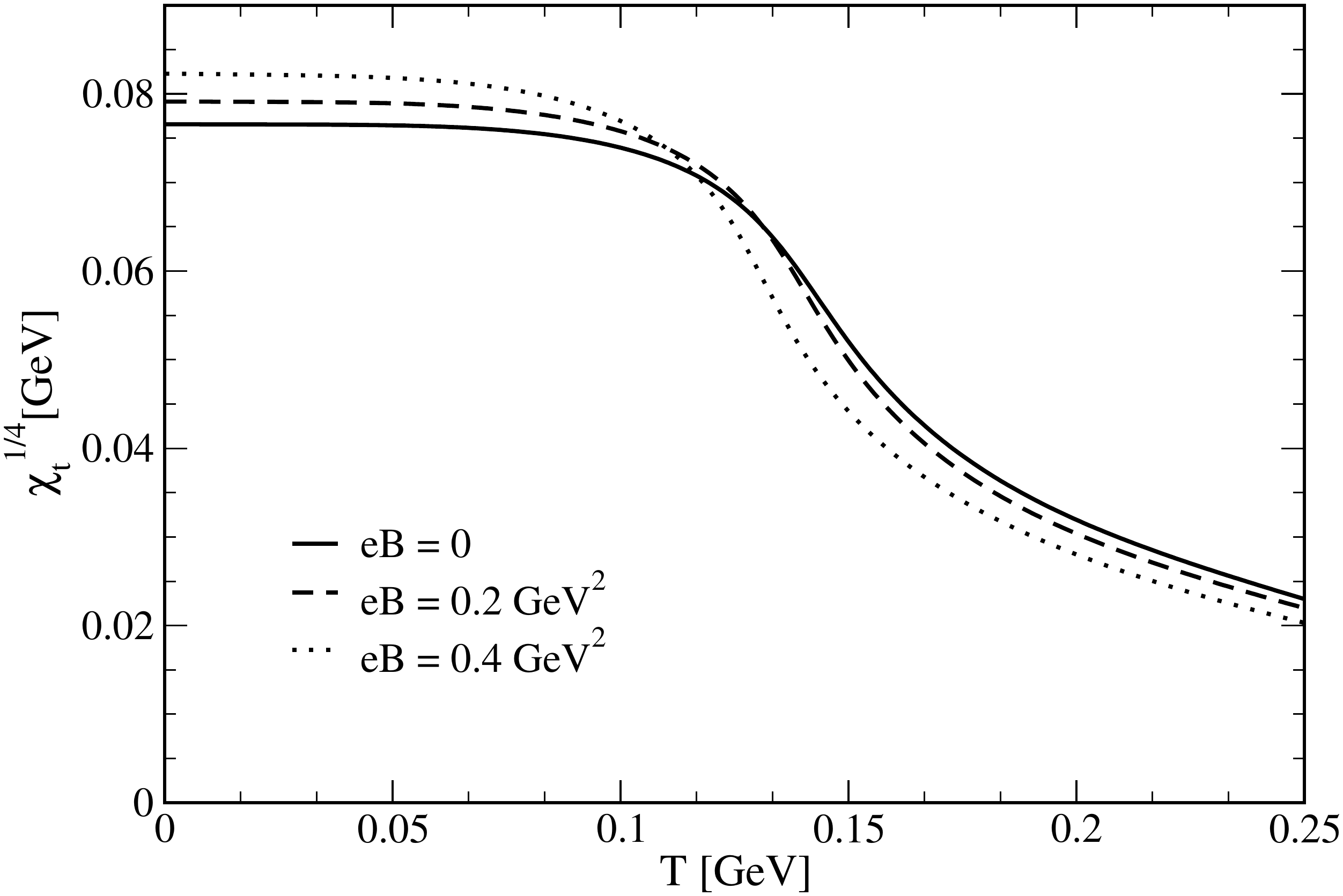}}
\caption{Same as in {}Fig.~\ref{ma_lambda_fixedG} but considering the
  fully $eB$ and $T$ dependent coupling constant $G_s(eB,T)$ for the
  axion mass ratio $m_a(T)/m_a$, panel (a), the  axion self-coupling
  ratio $\lambda_a(T)/\lambda_a$, panel (b) and for the topological
  susceptibility, panel (c).}
\label{ma_lambda_varyG}
\end{figure}
\end{center}

In {}Fig.~\ref{ma_lambda_varyG} we have shown the variation of scaled
axion mass, self-coupling and topological susceptibility with $T$ for
the fully magnetic field and temperature dependent coupling constant $G_s(eB,T)$. The MC
effect at lower temperature is not visible in
{}Figs.~\ref{ma_lambda_varyG}(a) and \ref{ma_lambda_varyG}(b) because
of the scaling used, but for both these cases the IMC is clearly
noticeable around $T_c$: The inflection points, indicating the
pseudo critical temperature location, is shifted towards lower
temperature with increasing magnetic fields. The kink-like feature in
the position of the pseudo critical temperature location appearing
in the axion self-coupling plot shows a similar behavior. In
{}Fig.~\ref{ma_lambda_varyG}(c) the topological susceptibility is also
shown in view of the fully $B$ and $T$ dependent coupling constant $G_s(eB,T)$,
which again shows both the MC and the IMC effects in different
temperature regimes with a crossover happening in between. The
quantitative difference between the zero temperature values of $\chi_{
  \rm{top}}$ shown in {}Fig.~\ref{ma_lambda_fixedG}(c) and
{}Fig.~\ref{ma_lambda_varyG}(c) is solely due to the difference in the
coupling constant $G_s$ parametrization, i.e., the difference between
the extrapolation of lattice fitted $G_s(eB,T)\big|_{T,eB=0}=4.6311$
GeV$^{-2}$ with the fixed $G_s = 5.022$ GeV$^{-2}$ case shown in
{}Fig.~\ref{ma_lambda_fixedG}(c).

\begin{center}
\begin{figure}
\includegraphics[width=8cm]{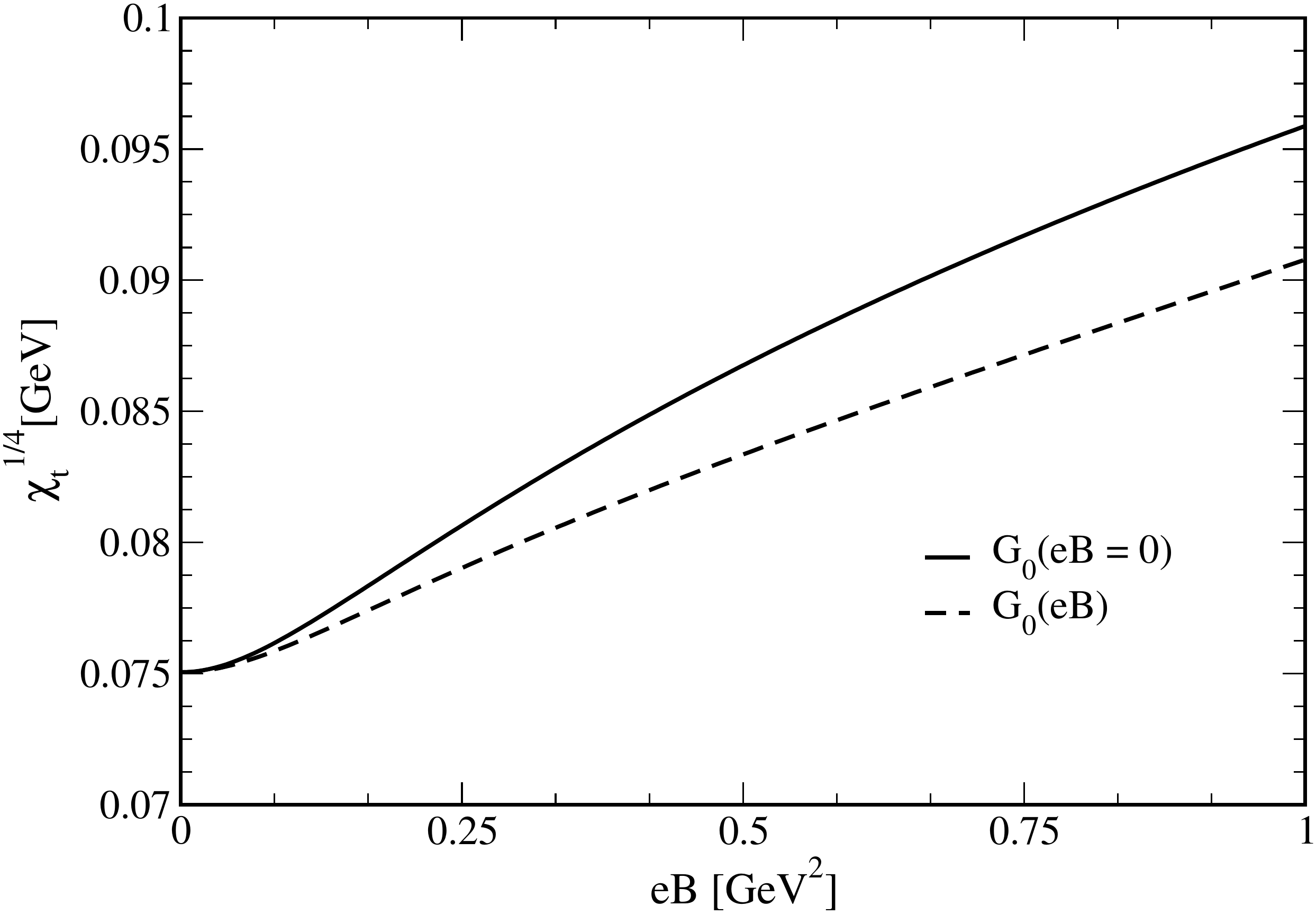} 
\caption{Variation of the topological susceptibility $\chi_{\rm{top}}$
  with respect to $eB$ in the zero temperature case. The solid curve
  is for the $G_0(eB=0)$ and  the dashed curve is when the fully
  magnetic field dependent coupling $G_0(eB)$ is considered.}
\label{topsus_T0_eB}
\end{figure}
\end{center}

{}Finally, as mentioned in Sec.~\ref{subsec2c}, at $T=0$ we can now
predict the behavior of the magnetic field dependence of the
topological susceptibility $\chi_{\rm{top}}$ using the magnetic field
dependent coupling $G_0(eB)$. This is explicitly shown in
{}Fig.~\ref{topsus_T0_eB},  where the variation for both $G_0(eB)$ and
when taking a fixed coupling constant, putting $B=0$ in
Eq.(\ref{T0_Gfit}), i.e., $G_0(eB=0) = 4.50373$
GeV$^{-2}$, are considered.

\begin{figure*}[!]
\begin{center}
\subfigure[ $\eta$ for $a/f_a = 0$.]{\includegraphics[width=8cm]{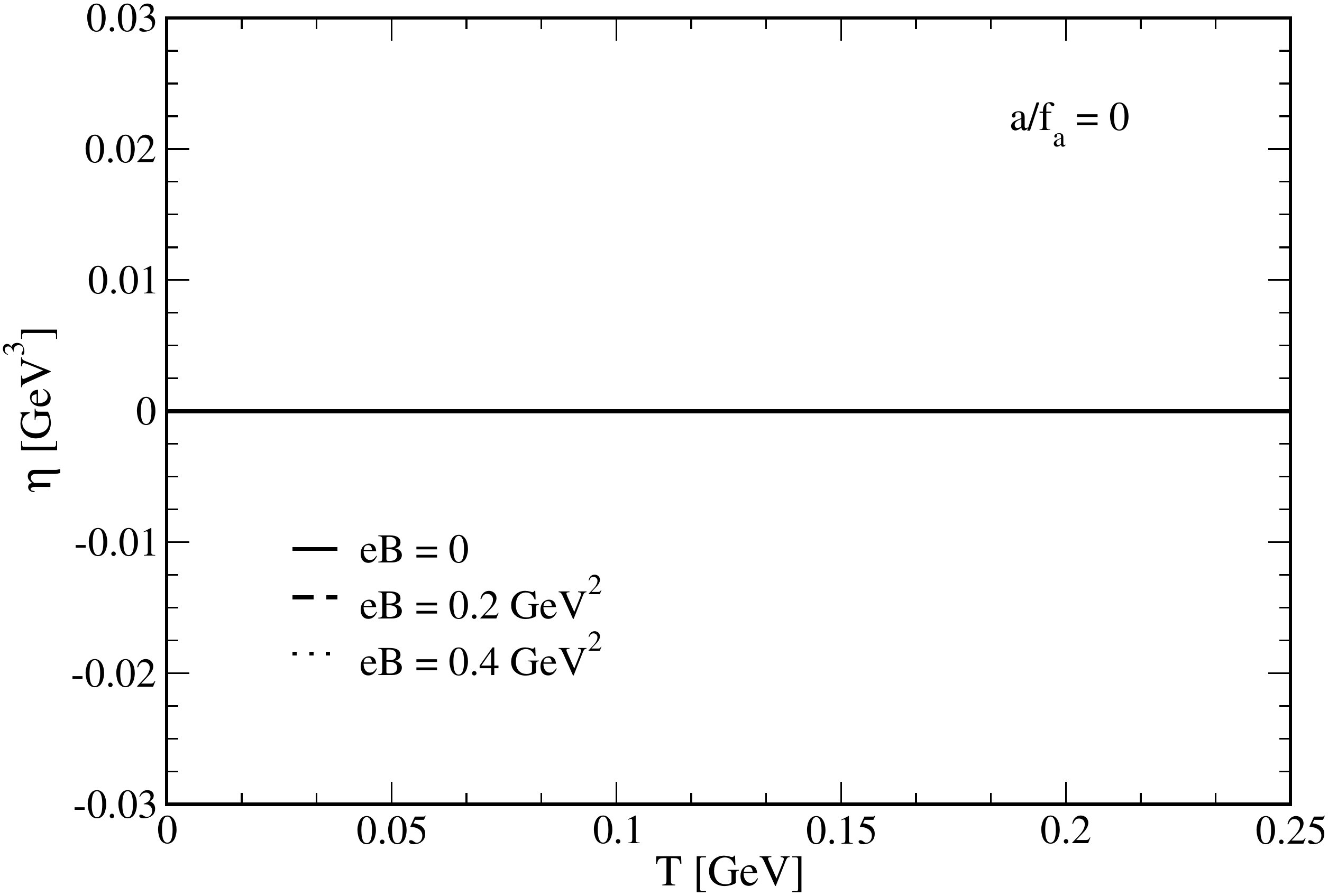}}
\subfigure[ $\eta$ for $a/f_a = \pi$.]{\includegraphics[width=8cm]{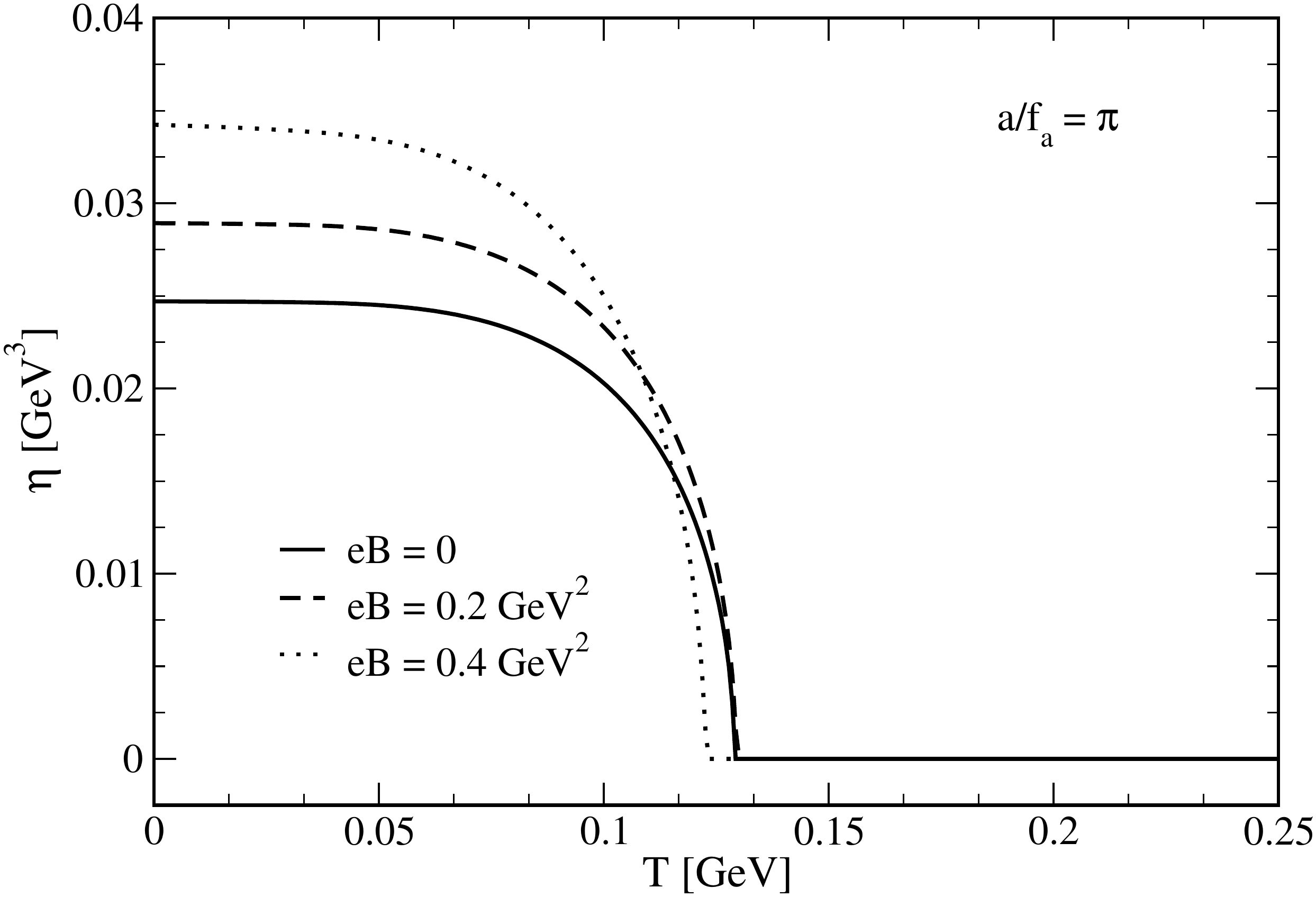}}
\subfigure[ $\sigma$ for $a/f_a = 0$.]{\includegraphics[width=8cm]{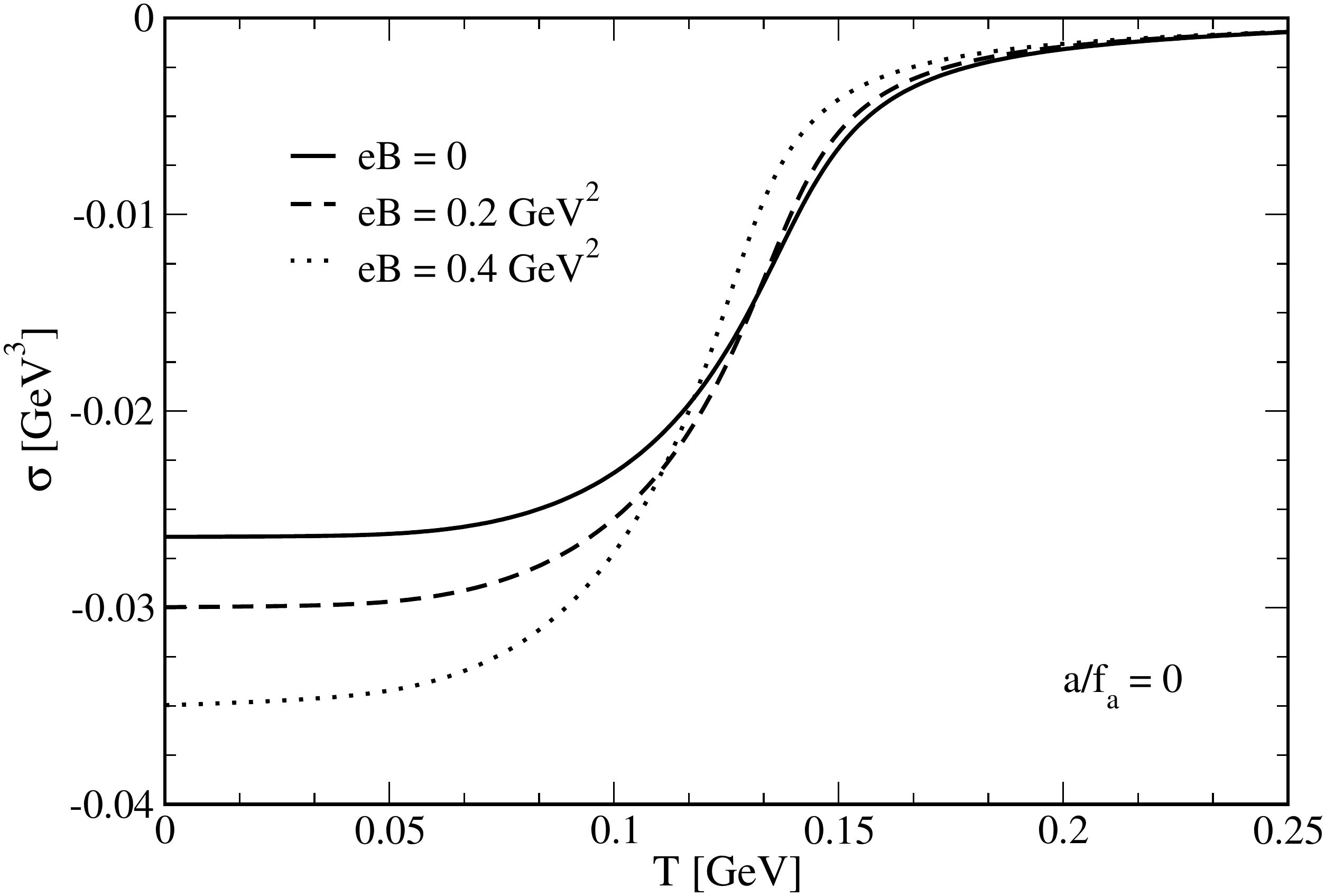}}
\subfigure[ $\sigma$ for $a/f_a = \pi$.]{\includegraphics[width=8cm]{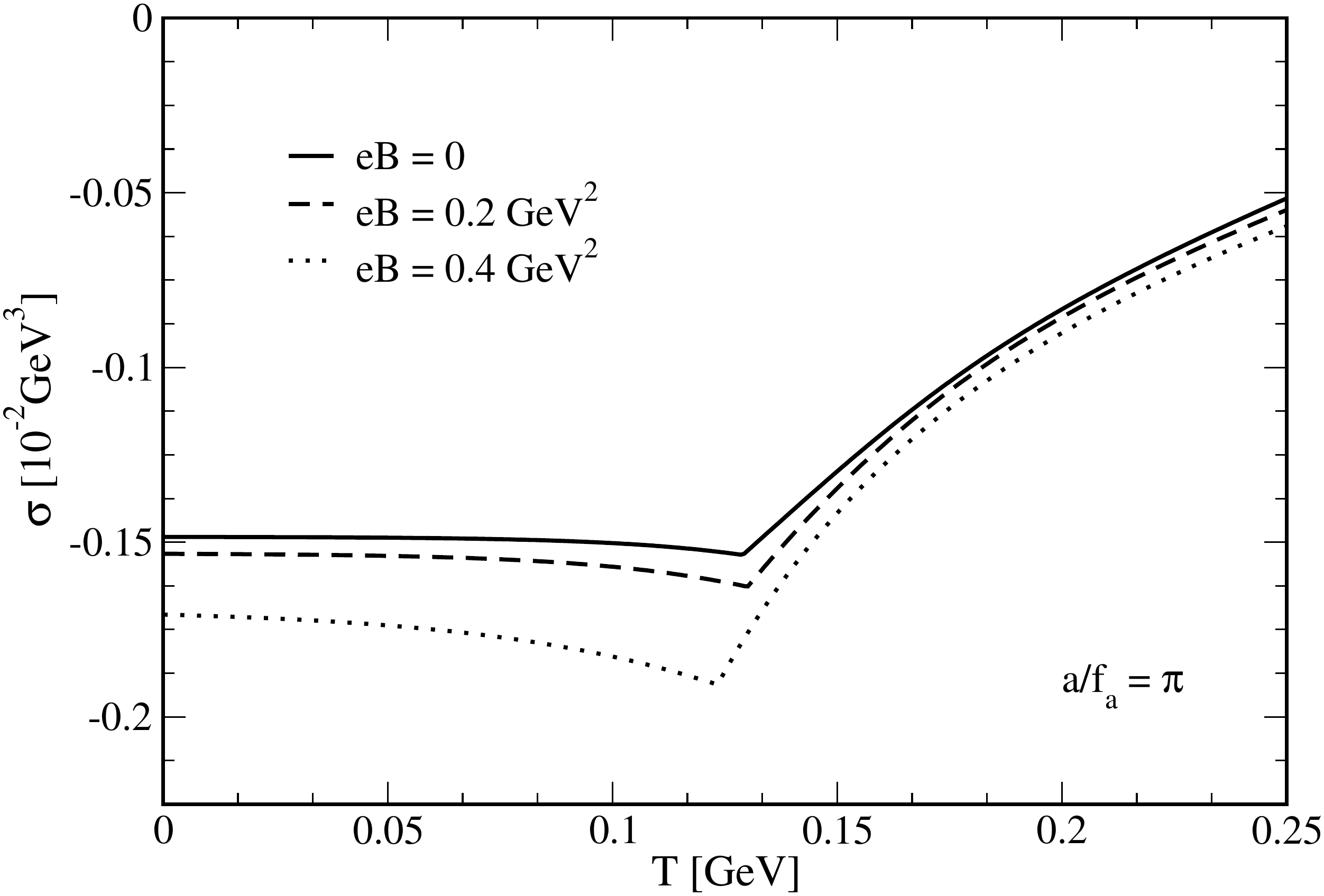}}
\caption{The temperature dependence for  the condensates $\eta$ and $\sigma$ for 
three different values of external magnetic fields with $G_s(eB,T)$. }
\label{condensates_vs_T}
\end{center}
\end{figure*}

As a final remark here, we should point out that as a central theme of our present work, 
let us discuss about the effect of the spontaneous CP violation observed in our results. 
In {}Fig.~\ref{condensates_vs_T}, we have shown the variations for both the $\sigma$ and $\eta$ 
condensates with respect to the temperature and for three different values of the external magnetic field. 
As a consequence of our discussions and results up to this point, we have only considered 
the running coupling $G_s(eB,T)$ when getting these results. Also, for illustration purposes, we have chosen two specific 
values for $a/f_a$, i.e., $0$ and $\pi$, which we believe should suffice for our discussion. 
The result shown in {}Fig.~\ref{condensates_vs_T}(a) illustrates that there is no $CP$ violation happening 
at $a/f_a=0$, irrespective of the increasing temperature and external magnetic field. 
This result is compatible with the well known Vafa-Witten theorem~\cite{Vafa:1984xg}. 
Now, the results displayed in {}Fig.~\ref{condensates_vs_T}(b) clearly show the $CP$ violating 
$\eta\neq 0$ phase for $a/f_a=\pi$ in the lower temperature region. 
The external magnetic field dependence of the critical temperature $T_c(eB)$ is also quite 
visible from {}Fig.~\ref{condensates_vs_T}(b), which gradually decreases with increasing 
$eB$ (also in agreement with the IMC phenomena). This result helps to show both the spontaneous $CP$ 
symmetry breaking at $T=0$, i.e., Dashen's phenomena, and the gradual restoration of the same at 
$T=T_c(eB)$. Then, in {}Figs.~\ref{condensates_vs_T}(c) and \ref{condensates_vs_T}(d), they show similar 
variations for the chiral condensate. They mimic the nature of the constituent quark mass shown in 
{}Fig.~\ref{eff_mass_varyG}(a). As a summary of the possible phases involved in our work, 
we can say that for $a/f_a = 2j\pi$ we have only the chiral symmetry breaking phase,
i.e., $\sigma\neq 0$ and $\eta=0$, which becomes restored at a temperature $T>0.2$ GeV,
i.e., leading to $\sigma=\eta=0$. But for $a/f_a= (2j+1)\pi$, apart from the chiral phase transition (a crossover), we have another second order phase transition at a magnetic field dependent critical temperature $T_c(eB)$, 
when the spontaneous $CP$ violating phase gets restored. For $a/f_a=\pi$, these three phases have explicitly been shown in the $T-c_{\rm min}$ plane in Fig.~\ref{ccrit_vs_T}.

\section{Conclusion}
\label{sec4}

In conclusion, we would like to convey that in the present work we
have studied, for the first time as far as it is of our knowledge, 
the thermodynamics
of QCD axions within the NJL model in a hot and magnetized medium by
explicitly emphasizing the effects of two of the most appealing
phenomena in this context, i.e., the magnetic catalysis and the inverse
magnetic catalysis effects. In the process, we have studied the different behaviors shown by the condensates for two different values of the scaled axion field, i.e. $a/f_a=0$ and $\pi$, consistent with the well known Vafa-Witten theorem and Dashen's phenomena respectively, leading to a discussion about the occurrence of different possible phases. Furthermore we have investigated the effect of the temperature and the external magnetic field explicitly on the measure of the spontaneous $CP$ violation, thereby showing the magnetic field dependence of the critical temperature for the spontaneous $CP$ symmetry restoration. Choosing the VEV of the axion field as a $CP$ violating term in the QCD action our results extend the previous results in the context of strong $CP$ violation~\cite{Frank:2003ve,Boomsma:2009eh,Boomsma:2009yk,Chatterjee:2014csa,Fukushima:2001hr} by incorporating Inverse Magnetic Catalysis effect in the medium using thermomagnetic field dependent coupling constant $G_s(eB,T)$. On top of that, the axion mass, self-coupling and topological susceptibility have their own significance regarding the study of cold dark matter, axion stars, the cooling anomaly problem in astrophysics, etc, and we have explored the effects of temperature and external
magnetic field on these quantities. Finally throughout this present work, we have also explored the importance of evaluating the differences in the results when different parametrizations are considered in the context of an effective model for QCD.  In this work, we have explicitly made this
study in the context of the two-flavor NJL model applied to the QCD
axion thermodynamics.  These results thereby strengthen the claim of
a strong dependence on parametrization within the NJL model.  

\section{Acknowledgements}

This work was partially supported by Conselho Nacional de
Desenvolvimento Cient\'{\i}fico e Tecnol\'ogico (CNPq)  under Grants
No. 304758/2017-5 (R.L.S.F.), No. 136071/2018-0 (B.S.L.) and No. 302545/2017-4 (R.O.R.),
Funda\c{c}\~ao Carlos Chagas Filho de Amparo \`a Pesquisa do Estado do
Rio de Janeiro (FAPERJ)  under Grant No.  E-26/202.892/2017 (R.O.R.)
and Coordena\c c\~ao de Aperfei\c coamento de Pessoal  de N\'ivel
Superior (CAPES) (A.B.) - Brasil (CAPES)- Finance Code 001.


\end{document}